\documentclass[%
 aip,
 floatfix,
rsi,%
 amsmath,amssymb,
preprint,%
]{revtex4-1}

\usepackage{graphicx}
\usepackage{dcolumn}
\usepackage{bm}

\begin{document}

\def \bc_figure_width{3.25in}
\preprint{AIP/123-QED}

\title[The aCORN Experiment]{aCORN: an experiment to measure the electron-antineutrino correlation coefficient in free neutron decay}

\author{B. Collett}
\email{bcollett@hamilton.edu.}
\affiliation{Physics Department, Hamilton College, Clinton, NY 13323, USA}

\author{F. Bateman}
\affiliation{National Institute of Standards and Technology, Gaithersburg, MD, 20899, USA}

\author{W. K. Bauder}
\altaffiliation[Current address: ]{Applied Research Laboratory, The Pennsylvania State University, State College, PA 16804, USA}
\affiliation{Physics Department, Hamilton College, Clinton, NY 13323, USA}

\author{J. Byrne}
\affiliation{University of Sussex, Brighton, Sussex, BA1 9QH, UK}

\author{W. A. Byron}
\affiliation{Department of Physics and Engineering Physics, Tulane University, New Orleans, LA 70118, USA}

\author{W. Chen}
\affiliation{University of Maryland, College Park, Maryland 20742, USA}
\affiliation{National Institute of Standards and Technology, Gaithersburg, MD, 20899, USA}

\author{G. Darius}
\affiliation{Department of Physics and Engineering Physics, Tulane University, New Orleans, LA 70118, USA}

\author{C. DeAngelis}
\affiliation{Department of Physics and Engineering Physics, Tulane University, New Orleans, LA 70118, USA}

\author{M. S. Dewey}
\affiliation{National Institute of Standards and Technology, Gaithersburg, MD, 20899, USA}

\author{T. R. Gentile}
\affiliation{National Institute of Standards and Technology, Gaithersburg, MD, 20899, USA}

\author{M. T. Hassan}
\affiliation{Department of Physics and Engineering Physics, Tulane University, New Orleans, LA 70118, USA}

\author{G. L. Jones}
\affiliation{Physics Department, Hamilton College, Clinton, NY 13323, USA}

\author{A. Komives}
\affiliation{Department of Physics and Astronomy, DePauw University, Greencastle, IN 46135, USA 
}

\author{A. Laptev}
\altaffiliation[Current address: ]{Los Alamos National Laboratory, Los Alamos, NM 87545, USA}
\affiliation{Department of Physics and Engineering Physics, Tulane University, New Orleans, LA 70118, USA}

\author{M. P. Mendenhall}
\affiliation{National Institute of Standards and Technology, Gaithersburg, MD, 20899, USA}

\author{J. S. Nico}
\affiliation{National Institute of Standards and Technology, Gaithersburg, MD, 20899, USA}

\author{G. Noid}
\affiliation{CEEM, Indiana University, Bloomington, IN 47408, USA}

\author{H. Park}
\altaffiliation[Visiting Scientist. Permanent address: ]{Korea Research Institute of Standards and Science, Daejeon, 34113, Korea}
\affiliation{National Institute of Standards and Technology, Gaithersburg, MD, 20899, USA}

\author{E. J. Stephenson}
\affiliation{CEEM, Indiana University, Bloomington, IN 47408, USA}

\author{I. Stern}
\affiliation{Department of Physics and Engineering Physics, Tulane University, New Orleans, LA 70118, USA}

\author{K. J. S. Stockton}
\altaffiliation[Current address: ]{Physics Department, University of Washington, Seattle, WA 98195}
\affiliation{Physics Department, Hamilton College, Clinton, NY 13323, USA}

\author{C. Trull}
\affiliation{Department of Physics and Engineering Physics, Tulane University, New Orleans, LA 70118, USA}

\author{F. E. Wietfeldt}
\affiliation{Department of Physics and Engineering Physics, Tulane University, New Orleans, LA 70118, USA}

\author{B. G. Yerozolimsky}
\altaffiliation{Deceased.}
\affiliation{Physics Department, Harvard University, Cambridge, MA 02139 USA}

\date{\today}

\begin{abstract}
We describe an apparatus used to measure the electron-antineutrino angular correlation coefficient in free neutron decay. The apparatus employs a novel measurement technique in which the angular correlation is converted into a proton time-of-flight asymmetry that is counted directly, avoiding the need for proton spectroscopy. Details of the method, apparatus, detectors, data acquisition, and data reduction scheme are presented, along with a discussion of the important systematic effects. 
\end{abstract}

\maketitle

\section{\label{sec:level1}Introduction}

The decay of the free neutron

\begin{equation}
\label{E:NDecay}
n \rightarrow p^+ + e^- + \bar{\nu}_e + 782\; keV
\end{equation}

\noindent provides a valuable laboratory in which to study the weak interaction, free from the added 
complexities of nuclear structure. The Standard Model relates the observable parameters, the 
neutron lifetime and the various angular correlation coefficients, to the underlying vector and axial 
vector coupling constants of the weak interaction, $g_V$ and $g_A$. In fact, there are more 
observables than independent components of $g_V$ and $g_A$, so that precision measurements of 
these observables overconstrain our knowledge of the values of the coupling constants, leading to 
the possibility of conflicts that would reveal the presence of physics beyond the Standard Model 
\cite{Abele}.

The probability of a neutron decay depends upon the neutron spin direction ($\hat{\sigma}$), the momentum and energy of the electron (${\vec p}_e$ and $E_e$), the energy released in the decay (Q), and the momentum and energy of the antineutrino (${\vec p}_\nu$  and $E_\nu$) according to the formula of Jackson, Treiman, and Wyld (JTW) \cite{JTW57}:

\begin{eqnarray}
\label{E:JTWeqn}
N & \propto & \frac{1}{\tau_n}E_e |{{\vec p}_e}| (Q-E_e)^2 \left[ 1 + a\frac{{{\vec p}_e}\cdot {{\vec p}_\nu}}{E_e E_\nu} \right. \nonumber \\
& & \left.  + \mbox{$\hat{\vec \sigma}$} \cdot \left( A\frac{{\vec p}_e}{E_e} + B\frac{{{\vec p}_\nu}}{E_\nu} + D\frac{({{\vec p}_e}\times {{\vec p}_\nu})}{E_e E_\nu} \right) \right],
\end{eqnarray}

\noindent where the dimensionless coefficients a, A, B, and D are the angular correlation coefficients and $\tau_n$ is the neutron lifetime.

The electron-antineutrino correlation coefficient ($a$-coefficient) governs the extent to which it is more probable for the electron and antineutrino to emerge traveling in correlated directions (${\vec p}_e \cdot {\vec p}_\nu > 0$) or in anticorrelated directions (${\vec p}_e \cdot {\vec p}_\nu < 0$). The value of the $a$-coefficient is closely related to the form of the weak decay Hamiltonian, which hypothetically could include scalar ($S$), tensor ($T$), axial vector ($A$), and vector ($V$) terms, categorized according to how each transforms under spatial rotations and reflections. The $S$, $V$ interactions create the beta electron and antineutrino in a spin singlet state (Fermi decay) while the $A$, $T$ interactions create them in a spin triplet (Gamow-Teller decay). This results in selection rules for the nuclear states in allowed beta decay: $\Delta J = 0$ for the Fermi case and $\Delta J = 0, \pm1$ for Gamow-Teller. Bloch and Miller \cite{BM35} pointed out that, additionally, the $S$, $T$ interactions require the beta electron and antineutrino to be emitted in the same helicity state, and the $V$, $A$ require opposite helicities, so each class of interaction leads to a different value of the $a$-coefficient: $-1$, $-\frac{1}{3}$, $+1$, $+\frac{1}{3}$ for pure $S$, $A$, $V$, $T$ decays, respectively. For mixed Fermi, Gamow-Teller decays, such as neutron decay, the $a$-coefficient obtains an intermediate value. This property of the $a$-coefficient was historically important in demonstrating the $V-A$ nature of the weak interaction \cite{Al59}. It remains a sensitive test of possible $S$, $T$ interactions due to new physics. Further elaboration on the physics of the
$a$-coefficient can be found in reference \onlinecite{Wie05c}.

At present, the $a$-coefficient is the least precisely known of the decay coefficients\cite{PDG}. In addition, all other current \cite{ASPECT1, ASPECT2} and previous \cite{PreviousLittle_a, Str78, Byr02} measurements of $a$ have relied on precision measurements of the proton recoil spectrum and thus share some of the same systematic limitations.
The apparatus described here and a future measurement \cite{NAB} rely on novel coincidence detection methods. Our measurement, proposed by Yerozolimsky and Mostovoy, \cite{Yer93, Bal94} offers different, and possibly smaller, systematic uncertainties from past measurements. 
It has been designed to hold all systematic effects to the level of \mbox{0.5 \%} of $a$ and thus to reach a final systematic uncertainty of approximately \mbox{1 \%} of $a$. The experiment took data for a period spanning 2 years on the NG-6 beamline at the National Institute of Standards and Technology (NIST) Center for Neutron Research (NCNR).

In this paper we present a detailed outline of the method (Section \ref{method}), and an overview of the apparatus (Section \ref{apparatus}). We describe each of the subsystems in detail starting with the neutron beam (Section \ref{beamline}), vacuum system (Section \ref{vacuum}), magnet system (Section \ref{BField}), collimator and electrostatic mirror insert (Section \ref{insert}), proton detector (Section \ref{PDet}), electron detector (Section \ref{betadet}), data collection (Section \ref{DAQ}), and data reduction  (Section \ref{reduction}). Finally, we describe the expected contributions to the systematic uncertainty and our approaches to estimating them in Section \ref{systematics}.

\section{\label{method}The $\textbf{\textsf{a}}$CORN method}

Previous determinations of the $a$-coefficient from the shape of the recoil proton spectrum encountered systematic limits in measuring the effect accurately at the 0.005 level (\mbox{5 \%} relative uncertainty in the $a$-coefficient). aCORN is based on a novel asymmetry method that does not require precise proton spectroscopy. It was first suggested about two decades ago by Yerozolimsky and Mostovoy \cite{Yer93, Bal94} and the proposed experiment was described in detail in a previous publication \cite{Wie05b}.

\begin{figure}[htbp]
\includegraphics[width=3.25in]{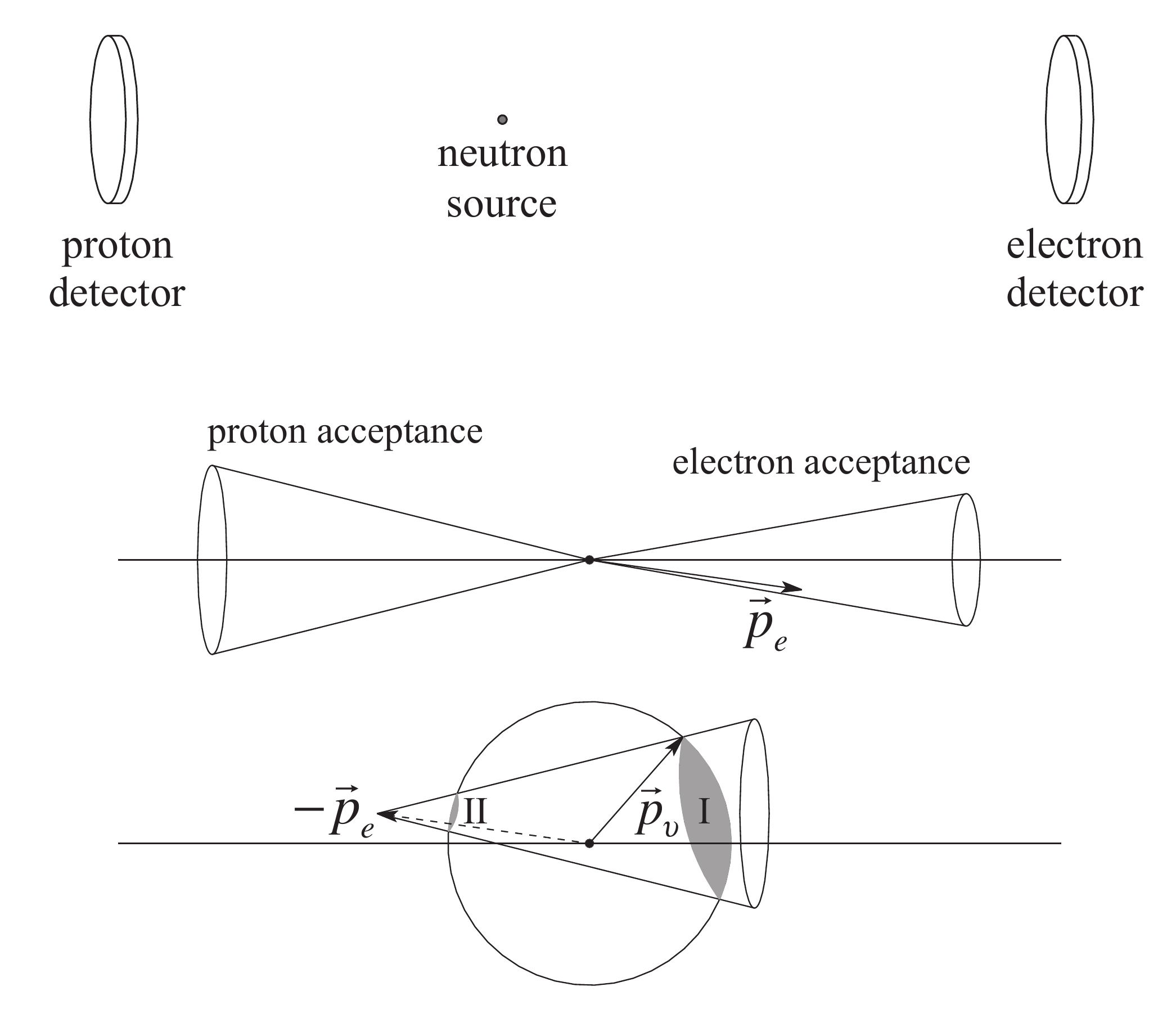}
\caption{\label{F:conesfig} Top: A simple scheme for detecting electron-proton coincidences from neutron beta decay. 
Middle: The corresponding momentum space diagram. The momentum acceptances for electrons and protons are cones. 
Bottom: A construction of the momentum acceptance for antineutrinos associated with the coincidence detection of a beta electron with momentum vector ${{\vec p}_e}$ and the recoil proton. Conservation of momentum and energy constrains the antineutrino momentum to the shaded regions I and II.}
\end{figure}

\par
To understand the experimental concept, let us first consider the simple configuration of Figure \ref{F:conesfig} (top). Proton and electron detectors are placed coaxially around a point source of decaying cold neutrons. The corresponding momentum space diagram is shown in Figure \ref{F:conesfig} (middle); the momentum acceptances of the beta electron and recoil proton are cones, which in general are not similar. The momentum vector for a particular detected electron is shown as ${{\vec p}_e}$. If the associated recoil proton momentum is inside the proton acceptance cone, then the electron-proton coincidence event is counted. The antineutrino is not detected, but because the neutrons decay effectively at rest,  the antineutrino momentum satisfies ${{\vec p}_{\nu}} = -{{\vec p}_e} - {{\vec p}_p}$
and the antineutrino momentum acceptance is the cone shown in Figure \ref{F:conesfig} (bottom), constructed by subtracting the proton cone from $-{{\vec p}_e}$. Since we have chosen a particular value of ${{\vec p}_e}$, the electron energy is fixed, and the proton kinetic energy is much less (smaller by a factor of about $10^{-3}$), so the antineutrino energy is determined to good approximation by the relation $E_{\nu} = Q_{\beta} - E_e$ and its momentum lies on a sphere as indicated. Whenever an electron-proton coincidence is detected, conservation of energy and momentum confines the antineutrino momentum to the intersection of the cone and the surface of this sphere: the two shaded regions labelled I and II. For region I events the electron and antineutrino momenta are correlated and for region II events they are anticorrelated. The two regions can be distinguished experimentally by proton time-of-flight (TOF); the protons corresponding to groups I and II form distinct TOF groups at each beta energy. The $a$-coefficient can then be obtained from the asymmetry in the number of region I and II events. However, even in the case $a = 0$, we see from the figure a large {\em intrinsic} asymmetry due to the difference in the effective antineutrino solid angles of the two regions; correlated antineutrinos ($\cos \theta_{e\nu} > 0$) are kinematically much more likely to produce a detected electron-proton coincidence. The small (order 10 \%) asymmetry due to the $a$-coefficient must be separated from this much larger intrinsic asymmetry. In practice the neutron source is an extended beam rather than a point source, so the intrinsic asymmetry depends on the decay position and a suitable convolution over the beam distribution must be made. This is not a favorable arrangement for a precision measurement of the a-coefficient.

\begin{figure}[htbp]
\begin{center}
\includegraphics[width = \bc_figure_width]{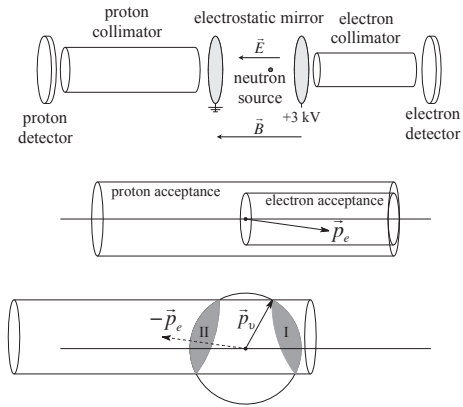}
\end{center}
\vspace{-0.25in}
\caption{\label{F:cylfig} Top: An illustration of the aCORN method. A uniform axial magnetic field $\vec{B}$ is applied throughout. Electron and proton collimators (depicted here as hollow cylinders) limit the transverse momentum accepted. An electrostatic mirror produces a uniform electric field  $\vec{E}$ near the decay region so that wrongly directed protons are turned around, hence any proton axial momentum is accepted. Middle: The corresponding momentum space diagram. The momentum acceptances for electrons and protons are cylinders. Bottom: A construction of the momentum acceptance for antineutrinos associated with the coincidence detection of a beta electron with momentum vector ${{\vec p}_e}$ and the recoil proton. Conservation of momentum and energy constrains the antineutrino momentum to the shaded regions I and II. In contrast to the scheme considered in Figure \ref{F:conesfig}, the solid angles of the two regions are equal.}
\end{figure}

\par
Now consider the scheme developed for the aCORN experiment and depicted in Figure \ref{F:cylfig} (top). Again, a point-like neutron decay source is viewed by coaxial proton and electron detectors, but now a uniform axial magnetic field ${\vec B}$ is applied and a set of proton and electron collimators is interposed. The magnetic field causes the charged particles to follow helical trajectories. This arrangement allows any proton (electron) whose transverse momentum is less than $eBr/2c$ to be detected, where $r$ is the proton (electron) collimator radius. The electron's axial momentum must be directed toward the electron detector to be counted. An electrostatic mirror produces a uniform axial electric field ${\vec E}$ near the decay region causing all wrongly directed protons to be reflected, so any value of the proton's axial momentum is accepted. The effect of the mirror field on the much more energetic beta electrons is negligible. The momentum acceptances for protons and electrons are bounded by the walls of the cylinders shown in the momentum space diagram, Figure \ref{F:cylfig} (middle). As before, we use ${{\vec p}_{\nu}} = -{{\vec p}_e} - {{\vec p}_p}$ and subtract the proton cylinder from $-{{\vec p}_e}$ to obtain the antineutrino momentum acceptance cylinder for a particular detected electron ${{\vec p}_e}$, Figure \ref{F:cylfig} (bottom). Again, to satisfy momentum and energy conservation, the antineutrino momentum is confined to the intersection of the acceptance cylinder and the surface of the sphere defined by $E_{\nu} = Q_{\beta} - E_e$, resulting in the shaded regions I and II.  By this construction we see that the solid angles for the two regions are equal, and the intrinsic asymmetry is (nearly) zero. This is the most important feature of the aCORN method. A measured asymmetry in the region I, II count rates is due to the $a$-coefficient alone. When the decay vertex is off-axis, as in the case of a beam source, the picture is somewhat more complicated because the momentum acceptance cylinders are elliptical rather than circular, but the construction and conclusions are similar: the region I and II solid angles are equal and the intrinsic asymmetry is zero. Another advantage of the aCORN method is that, in practice, the coincidence rate will be much higher compared to the scheme considered in Figure \ref{F:conesfig} because of the confining magnetic field and magnetic mirror.

\par
\begin{figure}[htbp]
\begin{center}
\includegraphics[width = \bc_figure_width]{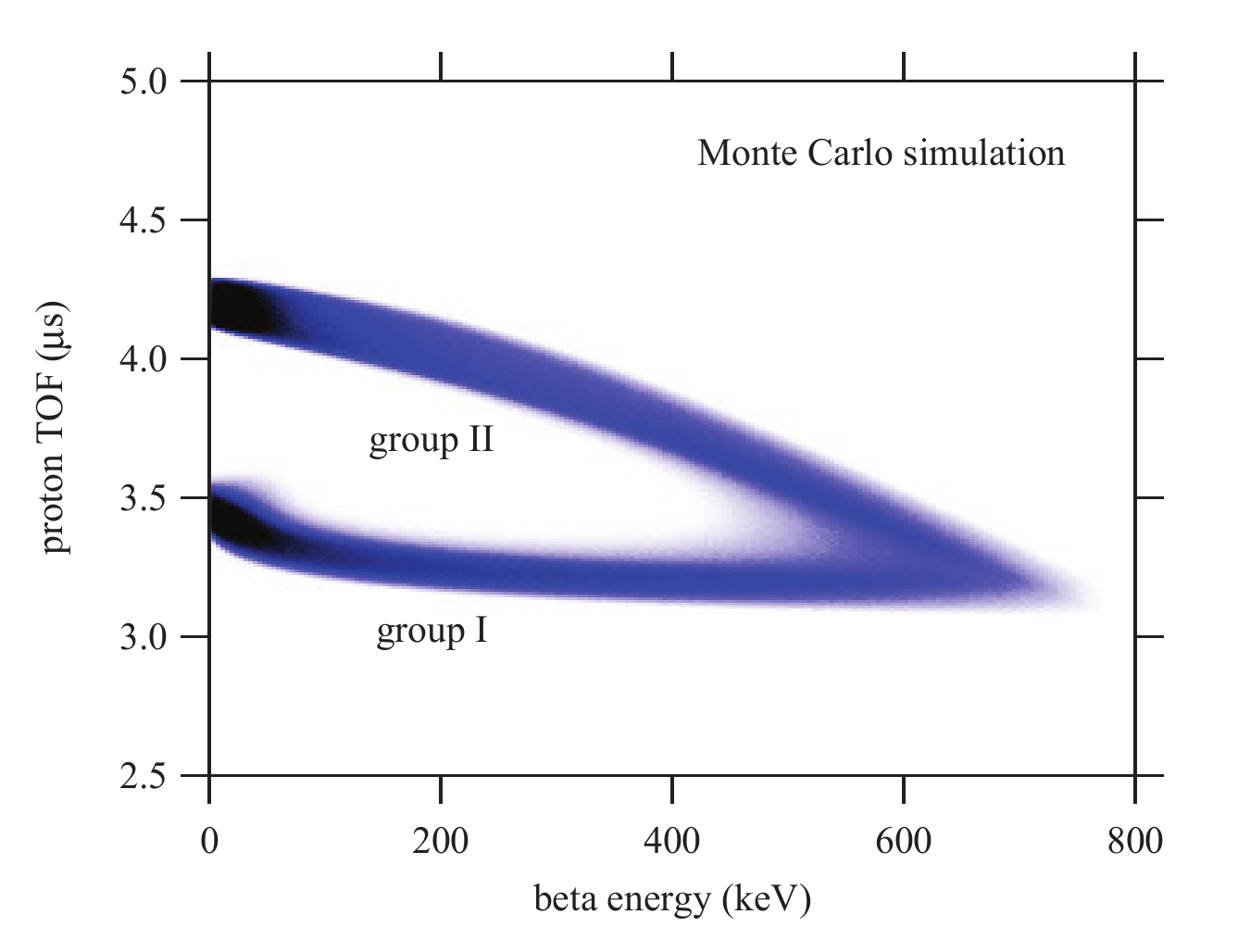}
\end{center}
\vspace{-0.25in}
\caption{\label{F:wbMC} A Monte Carlo simulation of aCORN data. The lower  branch of fast protons (group I) corresponds to events in the shaded region I in Figure \ref{F:cylfig}. The upper branch of slow protons (group II) corresponds to region II.}
\end{figure}

For each detected coincident event, the beta electron energy and electron-proton TOF are measured. The beta electrons are relativistic and are counted within a few nanoseconds of the decay, while the recoil protons are much slower and take several microseconds to reach the proton detector. Therefore, the TOF between the electron detection and the proton detection is a useful measure of the proton's initial axial momentum. A histogram of beta energy {\em vs.} TOF forms a characteristic wishbone shape, shown in Figure \ref{F:wbMC}. 

\par
The lower branch, fast protons, corresponds to the shaded region I in Figure \ref{F:cylfig} for which the electron and antineutrino momenta tend to be correlated. The upper branch, slow protons, corresponds to region II, where the momenta tend to be anticorrelated. The gap between the branches corresponds to the kinematically forbidden gap between regions I and II seen in Figure \ref{F:cylfig}. We obtain, after many decays, $N_{\rm I}$ events in group I (fast proton branch) and $N_{\rm II}$ events in group II (slow proton branch) for each electron energy slice.
Using equation \ref{E:JTWeqn} we have
\begin{equation}
N^{I(II)}(E) = F(E) \int\int \left( 1 + v a \cos\theta_{e\nu} \right) d\Omega_e\, d\Omega^{I(II)}_{\nu},
\end{equation}
where $F(E)$ is the beta energy spectrum, $v$ is the beta velocity (in units of $c$), $\cos\theta_{e\nu}$ is the cosine of the angle between the electron and antineutrino momenta, and $d\Omega_e$, $d\Omega^{I(II)}_{\nu}$ are elements of solid angle of the electron and antineutrino (group I, II) momenta. The integrals are taken over the momentum acceptances indicated in Figure \ref{F:cylfig}. Since the total solid angle products are equal for the two groups: 
$\Omega_e\,\Omega^I_{\nu}$ = $\Omega_e\,\Omega^{II}_{\nu}$, it is straightforward to show that the $a$-coefficient  is related to the experimental wishbone asymmetry $X(E)$:
\begin{eqnarray}
\label{E:asym}
X(E) & = & \frac{N^I(E) - N^{II}(E)}{N^I(E) + N^{II}(E)} \nonumber \\
& & = \frac{ \frac{1}{2} v a \left( \phi^I(E) - \phi^{II}(E) \right)}{1 + \frac{1}{2} v a \left( \phi^I(E) + \phi^{II}(E) \right)}.
\end{eqnarray}

\begin{figure}[htbp]
\begin{center}
\includegraphics[width = \bc_figure_width]{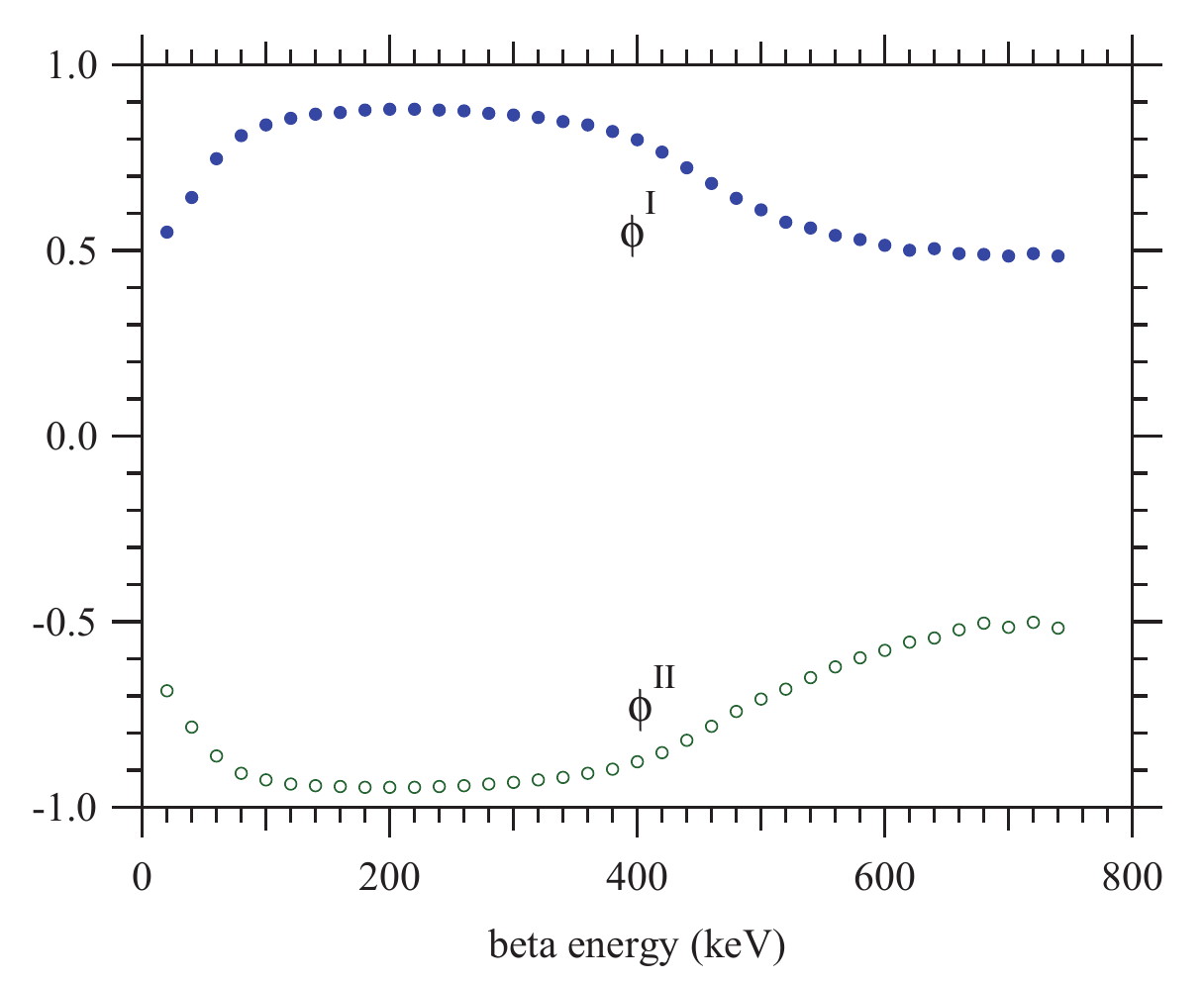}
\end{center}
\vspace{-0.25in}
\caption{\label{F:phis} A calculation of the aCORN geometric factors $\phi^I(E)$ and $\phi^{II}(E)$ (see equation \ref{E:phis}).}
\end{figure}

The parameters $\phi^I(E)$ and  $\phi^{II}(E)$ are defined by
\begin{eqnarray}
\label{E:phis}
\phi^I(E) = \frac{ \int d\Omega_e \int_I d\Omega_{\nu} \cos\theta_{e\nu} }{\Omega_e \Omega_{\nu}^I} \nonumber \\
\phi^{II}(E) = \frac{ \int d\Omega_e \int_{II} d\Omega_{\nu} \cos\theta_{e\nu} }{\Omega_e \Omega_{\nu}^{II}}.
\end{eqnarray}
Note that $\phi^I(E)$ and $\phi^{II}(E)$ can be understood as the average value of $\cos\theta_{e\nu}$ for each wishbone branch.  They are simply geometrical factors; they contain no physics and in particular they do not depend on the value of the $a$-coefficient. They depend on the transverse momentum acceptances of the proton and electron so they can be calculated from the known axial magnetic field and collimator geometries. A calculation of $\phi^I(E)$ and $\phi^{II}(E)$, using the actual parameters of the experiment, is shown in Figure \ref{F:phis}. 

\par
The second term in the denominator of equation \ref{E:asym} has a numerical value less than 0.005 in the energy range of interest (100--400 keV), so we can treat it as a second order correction and write:
\begin{equation}
\label{E:aEffective}
X(E) = a f_a(E)\left[ 1 + \delta_1(E) \right] + \delta_2(E)
\end{equation}
with
\begin{equation}
\label{E:faE}
f_a(E) = \frac{1}{2} v \left( \phi^I(E) - \phi^{II}(E) \right)
\end{equation}
and 
\begin{equation}
\delta_1(E) = -\frac{1}{2} v a \left( \phi^I(E) + \phi^{II}(E) \right).
\end{equation}
There is another correction that comes from neglecting the proton's kinetic energy in the momentum space discussion of Figure \ref{F:cylfig}. If we account for this energy, the antineutrino sphere is slightly oblong and the solid angles of groups I and II differ by approximately \mbox{0.1 \%}. The only non-negligible effect of this is to produce a small (about +0.0011) intrinsic asymmetry that is independent of the $a$-coefficient, represented by $\delta_2(E)$ in equation \ref{E:aEffective}. It is straightforward to compute this value to the needed precision using a Monte Carlo method.
\par
\begin{figure}[htbp]
\begin{center}
\includegraphics[width = \bc_figure_width]{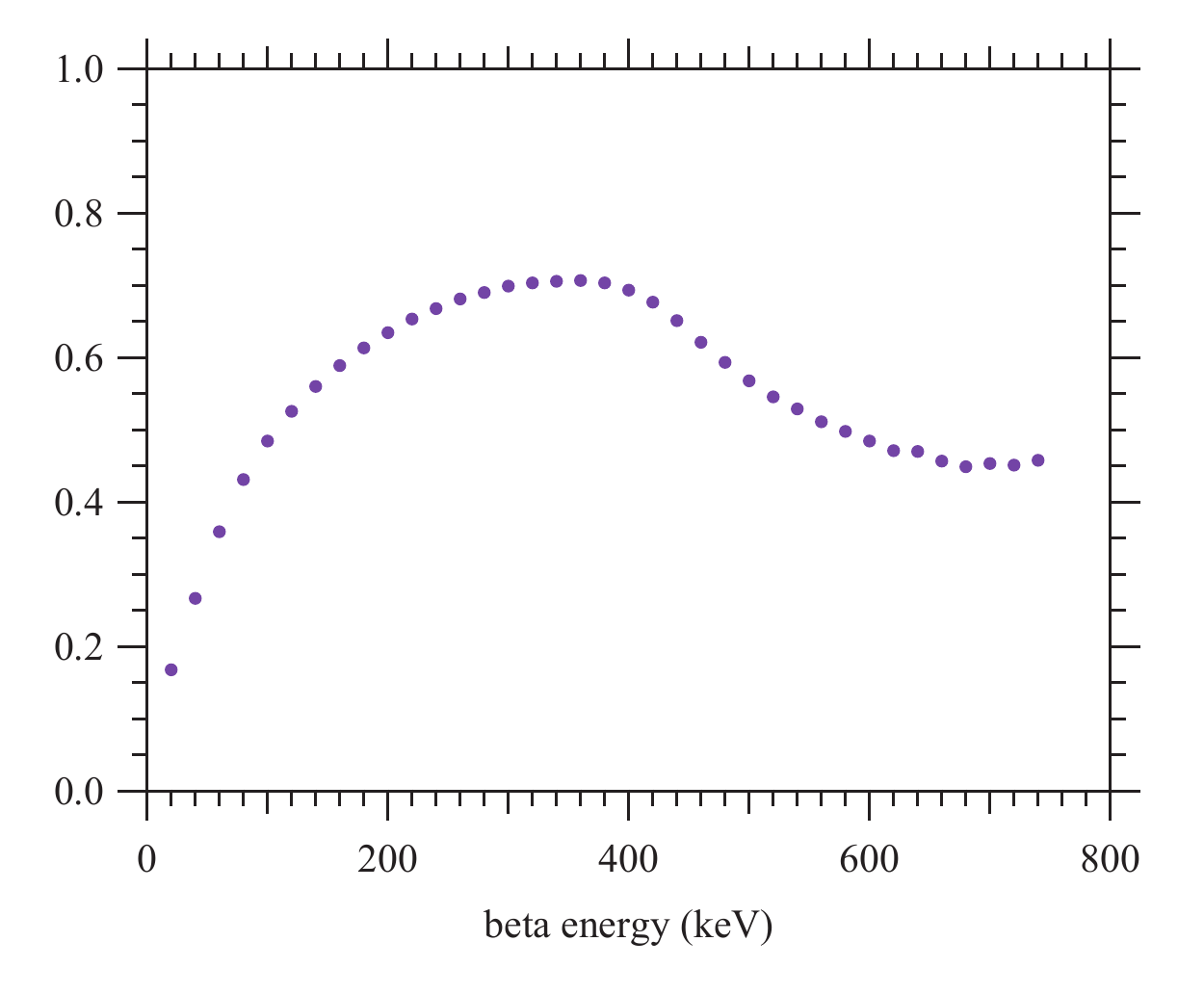}
\end{center}
\vspace{-0.25in}
\caption{\label{F:faE} The function $f_a(E)$ (see equations \ref{E:aEffective}, \ref{E:faE}), computed for aCORN using a Monte Carlo simulation.}
\end{figure}

Omitting the small corrections we see that $X(E) = a f_a(E)$; the experimental wishbone asymmetry is proportional to the $a$-coefficient and the geometric acceptance function $f_a(E)$, which in turn depends only on geometric factors and the electron mass so it can be precisely computed (Figure \ref{F:faE}).

\section{\label{apparatus}The Apparatus}

\par
The method described above makes a number of requirements for the apparatus. 
\begin{itemize}
\item There must be a way to get neutrons into and out of the decay region.
\item Since the charged decay products must travel significant distances, the experiment must be performed under high vacuum. 
\item There must be a detector that records the arrival time of the protons with nearly \mbox{100 \%} efficiency, independent of the proton phase space. 
\item There must be an energy resolving electron detector, which need not be perfectly
efficient or high resolution, but must be able to veto most events where the electron backscattered without depositing its full energy. 
\item The collimators and highly uniform axial magnetic field must be precisely aligned to provide the same transverse momentum acceptance for both wishbone branches.
\item There must be a highly uniform axial electric field in the neutron decay region. This must not extend into the proton collimator region, which must be free of electric field. For this experiment, the electric field is created by an arrangement of electrodes known as the {\it electrostatic mirror}.
\end{itemize}

\begin{figure}[htbp]
\begin{center}
\includegraphics[width = \bc_figure_width]{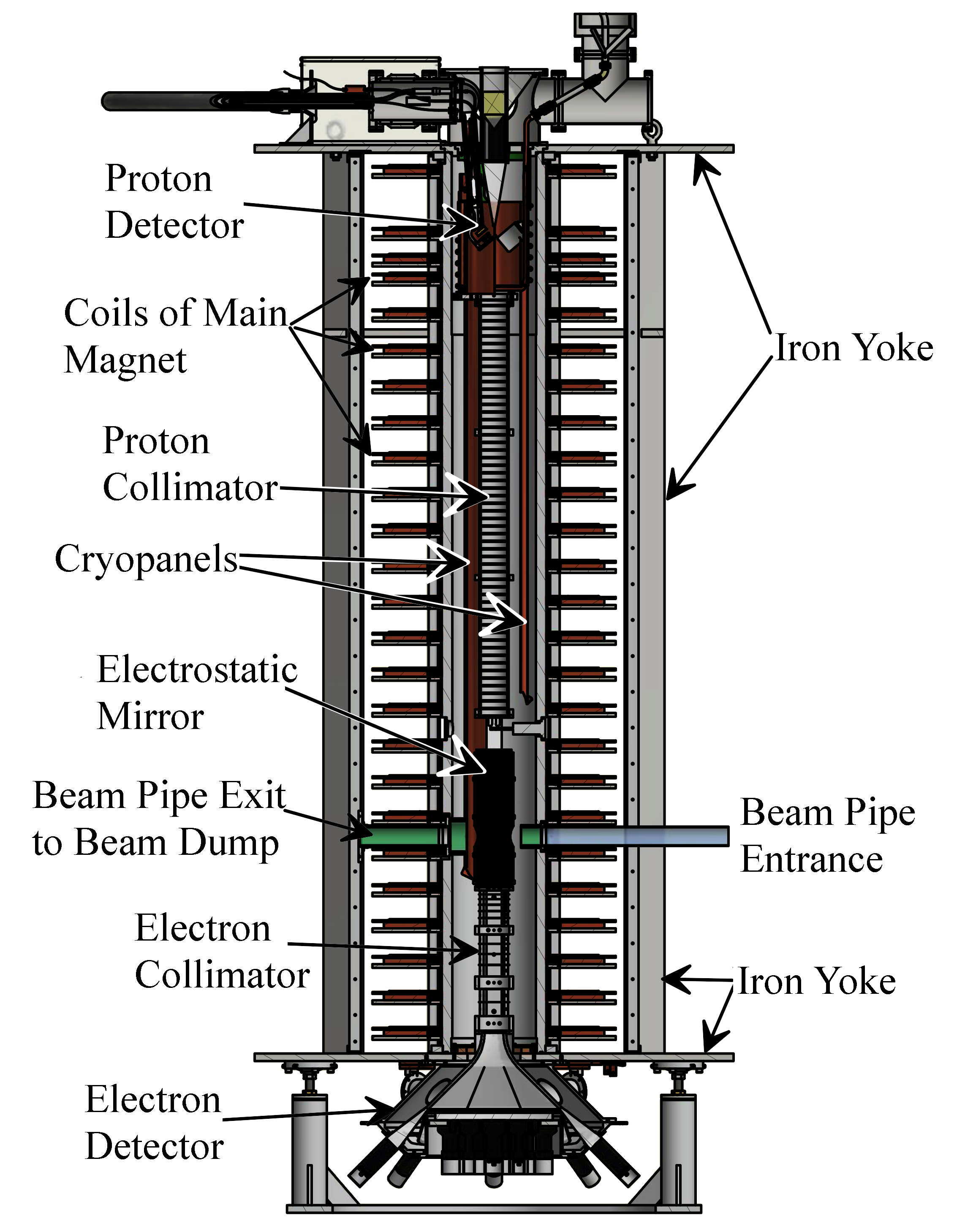}
\end{center}
\vspace{-0.25in}
\caption{\label{F:apparatus} Cross-section view of the aCORN experimental apparatus.}
\end{figure}

The resulting system is shown in schematic form  in Figure \ref{F:apparatus}. The magnetic field is generated by a segmented solenoid outside the vacuum vessel. The axis of the apparatus is defined by two bearings reproducibly mounted to the top and bottom of the vacuum vessel. In order to maintain the alignments between the collimators, the electric field, and the magnetic field, the collimators and mirror are held in place by a framework, known as the {\it{insert}}, that sits in the vacuum system. The individual components are first aligned within the insert and then the insert is aligned to the experimental axis. The individual subsystems and the alignment processes will be described in detail in the following sections.

\section{\label{beamline}Neutron Beamline}
\par
The aCORN experiment was installed and operated at the NG-6 fundamental neutron physics end position at the National Institute of Standards and Technology (NIST) Center for Neutron Research (NCNR) \cite{NCNR,Nic05}. A heavy water moderated, 20 MW research reactor produces thermal neutrons that are cooled by diffusion through a liquid hydrogen cold source at 18 K. A straight,  evacuated, $^{58}$Ni coated neutron guide, cross section 6 cm $\times$ 15 cm, efficiently transports the collimated cold neutron beam (average velocity approximately 800 m/s, average wavelength 0.5 nm) 68 m from the cold source to the experimental station. At the end of the neutron guide the beam passes through a thin aluminum-magnesium vacuum window followed by a liquid nitrogen cooled, 15 cm long, single-crystal bismuth filter that serves to remove fast neutrons and gamma rays from the beam. The neutron guide system and filter are surrounded by paraffin-filled steel shields to reduce radiation backgrounds.

\par
\begin{figure}[htbp]
\begin{center}
\includegraphics[width = 3.25in]{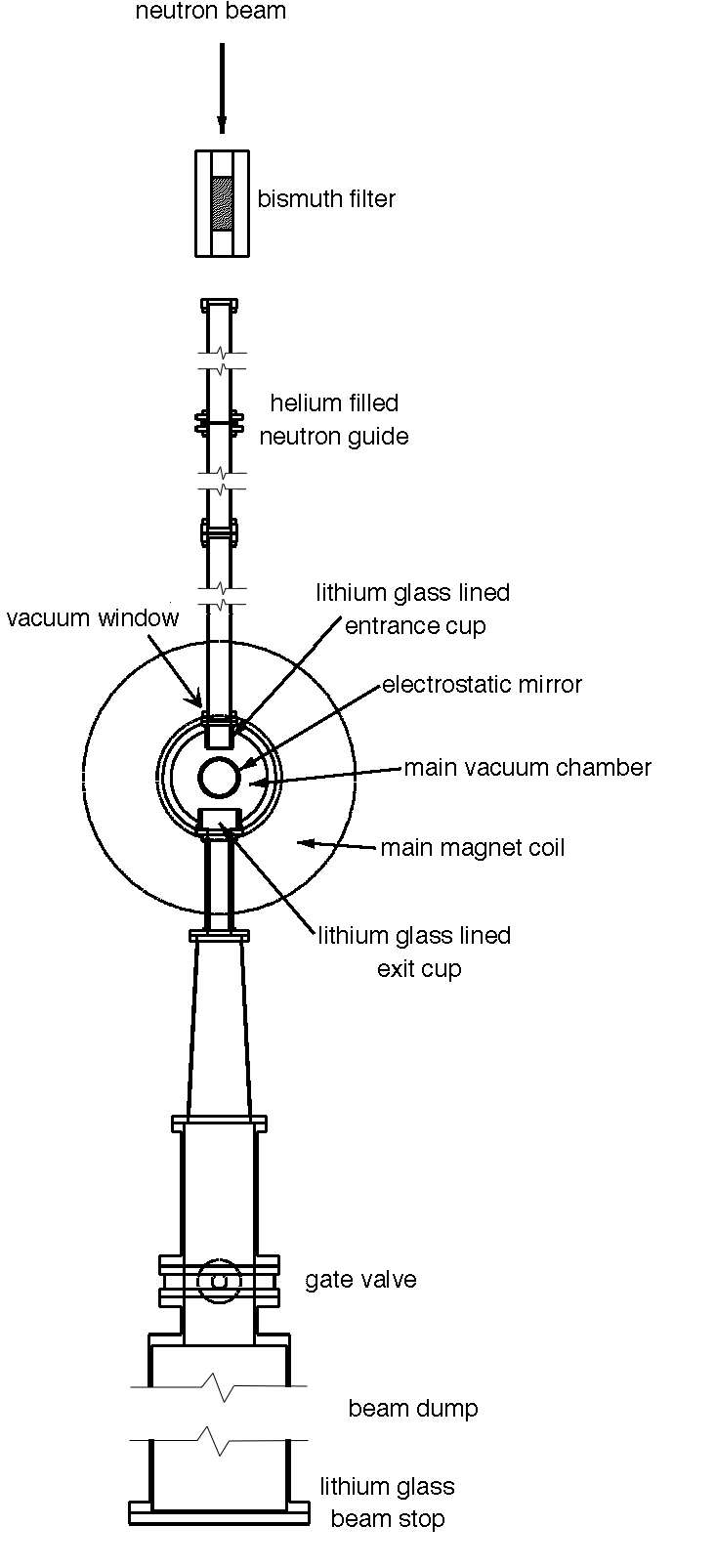}
\end{center}
\vspace{-0.25in}
\caption{\label{F:beamline} Section view from above of the experiment at beam level, illustrating the main components of the neutron transport.}
\end{figure}

The main components of the neutron transport are shown in Figure \ref{F:beamline}. After the filter, the neutron beam passed through a 3 m long borosilicate glass secondary guide, 6 cm in diameter. To improve neutron transmission, the neutron guide was filled with helium gas at a slight overpressure relative to atmosphere. End windows and the vacuum window were $\approx$0.1 mm thick beverage can aluminum. The last $\approx$0.5 m of guide penetrated into the gap between the sixth and seventh main magnet coils and ended at the main vacuum chamber. Within the chamber, beam entrance and exit cups were lined with $^6$Li-loaded glass to absorb scattered neutrons. During its passage through the vacuum chamber the full neutron beam passed twice through the wall of the electrostatic mirror, 4.4 $\mu$m copper electro-deposited on a 0.25 mm Teflon\cite{DISCLM} substrate. In order to maintain a highly uniform axial electric field inside the mirror, it was found necessary to pass the beam through the electrode system (see section \ref{sec:mirror} below), although neutron absorption in the copper contributed significantly to the radiation background. After exiting the main chamber, the neutron beam continued into the beam dump, a section of large diameter evacuated aluminum pipe culminating in the neutron beam stop, an aluminum flange covered with $^6$Li-loaded glass plates.

\par
\begin{figure}[htbp]
\begin{center}
\includegraphics[width = \bc_figure_width]{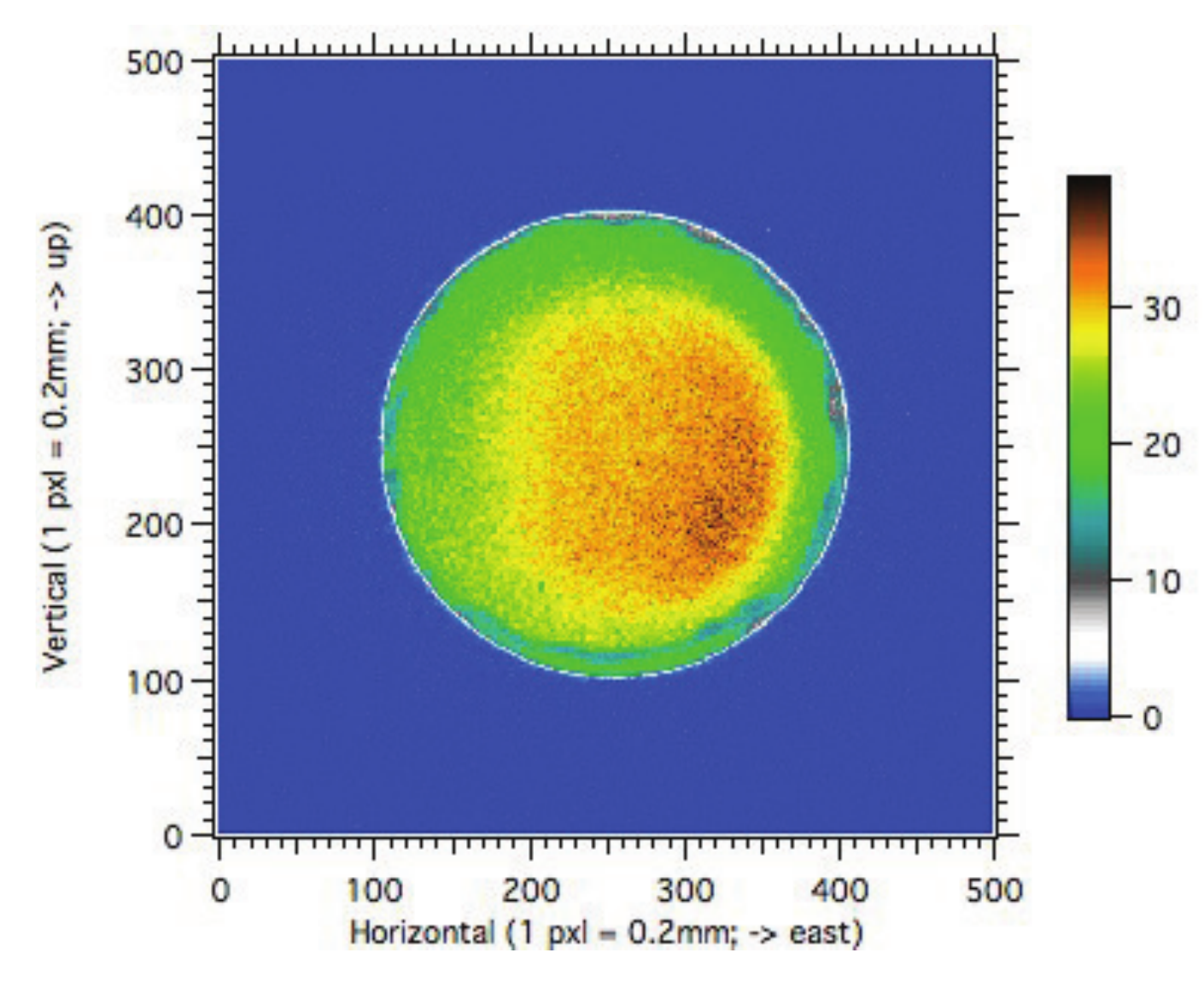}

\end{center}
\vspace{-0.25in}
\caption{\label{F:beamImg} An image of the aCORN neutron beam taken at the entrance window of the main vacuum chamber, made by exposing a thin dysprosium foil. The units of beam intensity are relative and the color scale has been chosen to enhance the contrast within the beam. }
\end{figure}

The neutron beam thermal equivalent flux (capture flux) at the entrance window to the main vacuum chamber was measured using a calibrated $^{235}$U fission chamber to be $6.9 \times 10^8$ cm$^{-2}$s$^{-1}$. Figure \ref{F:beamImg} shows an image of the neutron beam intensity at the entrance window, taken by the dysprosium foil exposure method \cite{Cho03}. A slight asymmetry in the beam intensity, enhanced by the false color, is evident in the figure. This is not unexpected and has no significant effect on the experiment.

\section{\label{vacuum}Vacuum System}
The main vacuum chamber of aCORN is an aluminum cylinder, 301 cm long, 28 cm inner diameter (ID), and 35 cm outer diameter (OD). The outside surface was turned to provide a highly cylindrical reference surface for the field coils. The top and bottom surfaces are o-ring sealed to mating regions machined into the upper and lower steel plates of the yoke (see section \ref{BField}). The electron detector is o-ring sealed to a similar surface machined into the bottom of the lower plate of the yoke and a vacuum cross similarly sealed above the top plate. One arm of the cross carries feedthroughs for electrical signals to the proton detector and the electrostatic mirror while another carries a liquid nitrogen feed for cooling. The beam dump is connected to the main vacuum system through a gate valve to the beam exit port in the side of the main vacuum chamber.

Pumping is provided by a 250 l/s turbomolecular pump on the electron spectrometer and a 370 l/s helium cryopump connected to the beam dump. In addition there are three liquid nitrogen cooled cryopanels that extend from the top of the vacuum chamber down to the bottom of the electrostatic mirror (Figure \ref{F:apparatus}), which provide additional high conductance pumping of water and volatiles. The pressure is monitored in three locations, near the cryopump, in the cross at the top of the apparatus, and through a side port just above the top of the electrostatic mirror. In normal operation the pressure at the top of the mirror is about $8\times10^{-5}$ Pa ($6\times10^{-7}$ Torr). A small amount of data were also collected at deliberately higher pressures, to study systematic effects due to residual gas. The turbo pump on the beta spectrometer was turned off, producing a pressure of approximately $3\times10^{-3}$ Pa ($2\times10^{-5}$ Torr), about 40 times the normal pressure.

\section{\label{BField}Magnetic Field}
The  magnetic field must transport protons from the electrostatic mirror to the proton detector in such a way that the transverse momentum acceptance of both proton groups is equal. We are most sensitive to the uniformity and alignment throughout the regions of the electrostatic mirror and the proton collimator. The requirement to transport the neutron beam in and out through the sides of the system that generates this field ruled out a single large solenoid. The field must also fall rapidly to a very low value at both the electron and proton detectors. The electron spectrometer (section \ref{betadet}) relies on a rapid decrease in the magnetic field for suppression of backscattered electrons. The proton detector operates at -28 kV and the combination of high electric fields and a large magnetic field can lead to high voltage instability. Finally, the orientation of the field must be reversible to check for effects caused by residual polarization of the neutron beam. 

The magnetic field is generated by a segmented solenoid consisting of 24 identical pancake coils. Each coil is wound from 121 turns of 2 cm $\times$ 0.1 cm copper tape on a form with an interior diameter of 44.5 cm and an exterior diameter of 78.8 cm attached to a 1.27 cm thick Al disk. Cooling is provided by a 0.6 cm thick copper plate with almost two turns of 0.6 cm copper tubing brazed into grooves on the top surface to carry cooling water. The complete coil assembly sits inside a magnetic flux return yoke made up from 2.54 cm thick, 160 cm diameter top and bottom plates held together with four vertical columns, 3.5 cm x 14 cm in cross-section and 301 cm long, all fabricated from 1008 carbon steel. There is an 18 cm diameter hole in the center of the lower plate to allow electrons to pass through to the electron detector and a 33 cm diameter hole in the top plate. In addition to supporting the system, the steel yoke serves as a flux return for the magnetic field, improving the field uniformity at the bottom and limiting the fringe field seen by nearby experiments.

The Al disk of the lowest coil sits directly on the bottom steel plate and then the next 21 coils are supported by brackets, so that the coils are all equally spaced, 12 cm from center to center leaving an 8 cm gap between coils. The individual coils slide tightly on the Al vacuum chamber, which helps center them on the experimental axis. Coils 23 and 24 are stacked directly on top of coil 22 to create an end-compensation and to reduce the magnetic field in the region of the proton detector. A twenty fifth coil was left unconnected and used as a spacer in the region of the transition to low field. The 24 active main coils are wired in series and carry a current of 28.17 A at a total of 140 V. Each coil dissipates about 170 W of power, which is easily removed by the water cooling.

In addition to the main coil, each of the forms also contains a small axial trim coil. Each trim coil is wound from 115 turns of 1 mm diameter  Cu wire filling a 2 cm tall space with an inner diameter 40.6 cm and outer diameter of 41.6 cm. Together, the trim coils help compensate for the end effects at top and bottom. The bottom steel yoke plate has additional coils placed on its upper and lower surfaces. On the upper surface, inside the vacuum vessel and separated from the yoke by a 2 cm thick Al flange, there is a coil wound from 69 turns of 2.3 mm square-section wire with an inner diameter of 18.3 cm and an outer diameter of 23.8 cm. This coil provides a field that mitigates the effect of the 18 cm diameter hole on the field inside the yoke and is cooled by contact with the Al flange. On the lower surface of the iron plate there is a 24 turn coil, 50 cm inner diameter, 60 cm outer diameter, and a thickness of 5 cm, that can boost the field just below the yoke. The additional field prevents the electrons emerging through the bottom plate of the yoke from spreading out too fast as the field below the plate falls to zero. This ensures that electrons accepted by the collimator reach the active area of the electron energy detector.

Transverse trim coils are used to cancel external horizontal fields in two perpendicular directions. A large pair of rectangular coils provides an approximately uniform field in each direction, while 24 pairs of rectangular trim coils in each direction are used to eliminate local transverse fields. 
These 12 cm tall coils are wound from 22 gauge wire on a frame just outside the main coils, with pairs separated by the 35 cm diameter of the vacuum vessel. The number of turns varies from 5 to 20 depending on the strength of correction required at each position. The axial and transverse trim coils are energized by individual low voltage/moderate current outputs from a computer-controlled multi-channel power supply, designed and fabricated for this experiment.

Numerical simulations have shown that a uniform $10^{-4}$ radian (0.1 mrad) misalignment of the magnetic field relative to the proton collimator axis will cause a false asymmetry equivalent to \mbox{0.5 \%} of $a$, the target limit for systematic effects. To meet our systematic uncertainty goal, in addition to requiring the magnetic field axis to be aligned to the proton collimator axis to within 0.1 mrad, we must have axial gradients less than 2 $\mu$T per cm, and know the absolute value of the magnetic field to within 200 $\mu$T.  The primary axis of the experiment is defined by two bearings \cite{Bearings,DISCLM}, each rigidly mounted to the flux return yoke and registered by dowel pins. With the insert removed, we measure the uniformity of the field and its alignment to this axis using a set of three 3-axis Hall probes \cite{SyprisHallProbe} within a temperature-controlled carriage that travels inside a square-section Al tube. A servo-motor system moves the carriage vertically within the tube under computer control.  The tube is mounted on the bearings at the top and bottom of the aCORN vacuum vessel and rotated by a second servo-motor system, also under computer control. 

During mapping, the carriage is moved vertically to positions every 2 cm along the axis. At each point data are collected from all three axes of the central probe at rotational positions from $0^\circ$ to $360^\circ$  at $30^\circ$  intervals. Rotation about the known axis set by the bearings allows the unknown (but fixed) misalignments of the probe axes to be distinguished from the actual variation of the field. The relative position of the servo-motor assembly is repeatable to a fraction of a millimeter and the absolute vertical position is calibrated by running high current through a select group of trim coils.  Since the trim coils are mechanically fixed and each creates a field with a well-known shape that has a maximum at its geometric center, the positions of these maxima provide reference points for the position calibration.

\begin{figure}[htbp]
\begin{center}
\includegraphics[width = \bc_figure_width]{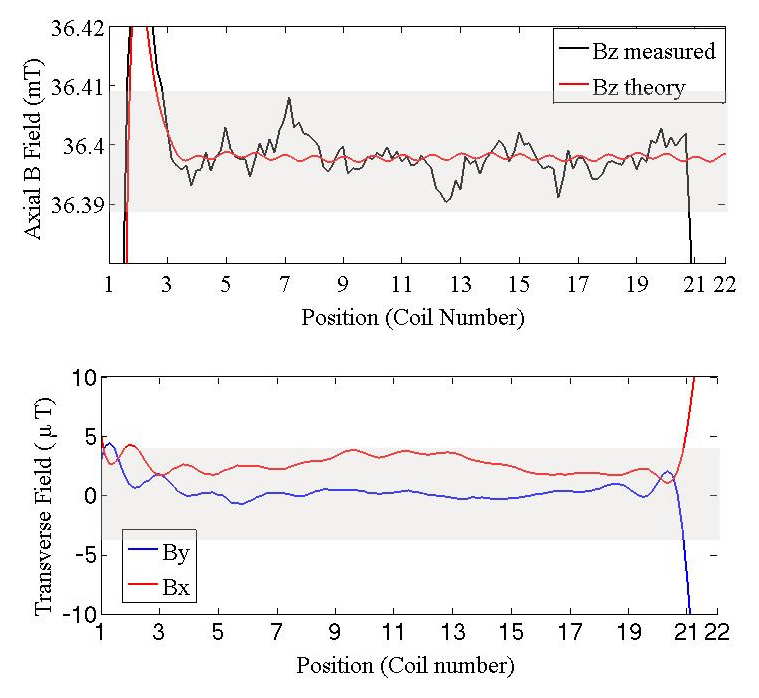}
\end{center}
\vspace{-0.25in}
\caption{\label{F:Plot} (Top) Map of axial magnetic field.
The red line is the ideal field from a computer model and the black is the measured field. The goal is a field that is flat within 0.2 mT (gray band) over the region from coil 6 to coil 21.\\
(Bottom) Map of the transverse components of the magnetic field. These components should average less than 4 $\mu$T (gray band) over the region from coil 6 to coil 21.}
\end{figure}

The magnetic field is mapped and trimmed using the mapper system and a computer model of the full set of trim coils. Typically, between 3 and 4 iterations of mapping and trimming are needed to converge on a set of trim coil currents that satisfies our requirements (Figure \ref{F:Plot}).

The 3-axis Hall probes used in the mapper system worked very reliably for relative field shape measurements, but their calibration was not sufficient to determine the absolute magnetic field. The absolute field strength is measured inside the proton collimator using $^3$He nuclear magnetic resonance (NMR). A 2.5 cm diameter cell containing  $^3$He, N$_2$, and Rb at a pressure of $3\times10^5$ Pa was polarized by spin-exchange optical pumping, then lowered into the proton collimator region of the apparatus.  Free induction decay NMR was used to measure the Larmor frequency of the $^3$He nuclei and thus the absolute magnetic field \cite{He3}.

Because both the mapper and the NMR field probes require the mapping system to be at atmospheric pressure, they can only be used when the experiment is not operating. During operation the magnetic field is monitored with a 3-axis fluxgate magnetometer inserted into fittings rigidly attached to several points on the aCORN superstructure. These monitor the fringe field outside the magnet where it is within the 1 mT range of the fluxgate. So long as the fringe field is stable then the field inside aCORN must also be stable. Under operating conditions the fluxgate sits near the beam entrance  and readings are recorded and logged every few minutes. A few times per week the flux gate was moved and a measurement was made at each of the fittings. The fluxgate measurements were stable to within 10 $\mu$T. In addition, the current in the main magnet is monitored using a precision shunt resistor in series with the axial coils.

\section{\label{insert}Insert}
\begin{figure}[htbp]
\begin{center}
\includegraphics[width = \bc_figure_width]{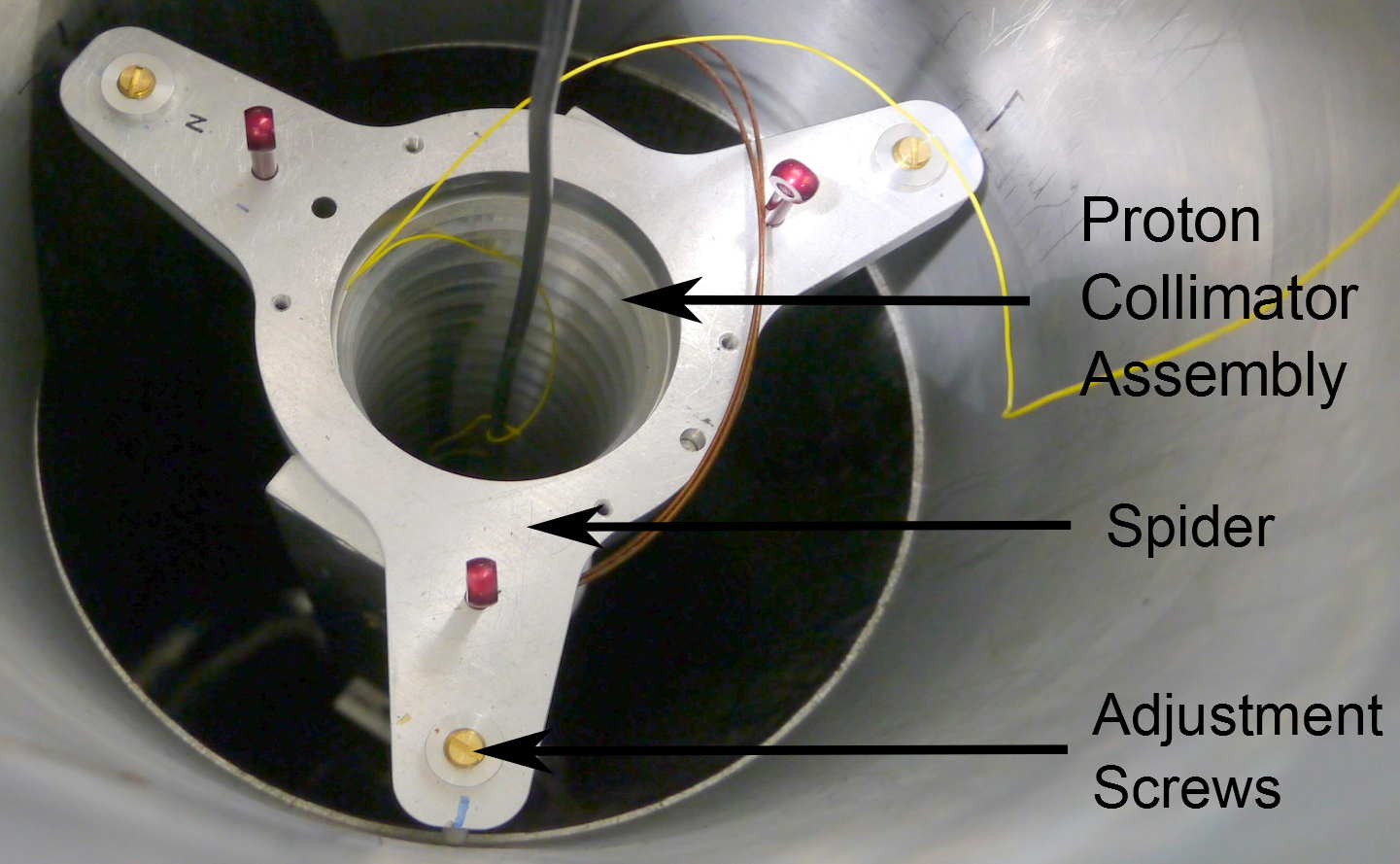}
\end{center}
\vspace{-0.25in}
\caption{\label{F:Insert} View down the top of the vacuum vessel showing the three-pronged spider and the adjustment screws that position it within the vacuum vessel. The top few rings of the proton collimator are also visible through the central aperture. The wires entering the collimator go to the NMR probe and were not present in normal operation.}
\end{figure}

The electron collimator at the bottom, the electrostatic mirror at beam height, and the proton collimator near the top are all attached to a rigid framework, the insert, that maintains their relative alignment and that supports them within the aCORN vacuum vessel. The insert sits in a collar at the bottom of the vacuum chamber and is held in place at its upper end by a three-pronged spider (Figure \ref{F:Insert}), which mates to the upper part of the inner wall of the vacuum chamber.  The arms of the spider end in small rods whose positions can be adjusted with screws. Thus the precise position of the top center of the insert can be adjusted {\it in situ}. This section will first describe the individual assemblies in some detail and will then describe the procedures used to guarantee that each assembly is correctly aligned to the experimental axis.

\subsection{Electron Collimator}
The geometry factors of equation \ref{E:phis} can be calculated for any well defined geometry. A Monte Carlo simulation was used to balance the increase in count rate that comes from a wider transverse momentum acceptance against the loss in useable energy range that this brings, as the two arms of the wishbone broaden and merge at a lower energy.  A 5.5 cm diameter cylindrical electron collimator and 8.0 cm diameter proton collimator were found to be optimal. A 782 keV electron with no axial momentum has a transverse momentum of 1188 keV/c, corresponding to a cyclotron radius of 10.9 cm in a 36.4 mT magnetic field.  A 5.5 cm diameter cylindrical collimator limits the maximum transverse momentum to 300 keV/c but is typically much more restrictive, depending on the radial position of the decay.  The Monte Carlo model predicts that  \mbox{1.8 \%} of the beta electrons from decays within the decay volume of the beam will be transmitted to the electron detector.

A cylindrical collimator as depicted in Figure \ref{F:cylfig} would be ideal if not for the large surface area it presents for scattering electrons, which would then be detected with the wrong energy. Instead we used a series of thin, circular apertures in an arrangement optimized using the PENELOPE\cite{penelope} electron simulation package  to minimize the probability of scattered electrons to reach the beta spectrometer.

\begin{figure}[htbp]
\begin{center}
\includegraphics[width = \bc_figure_width]{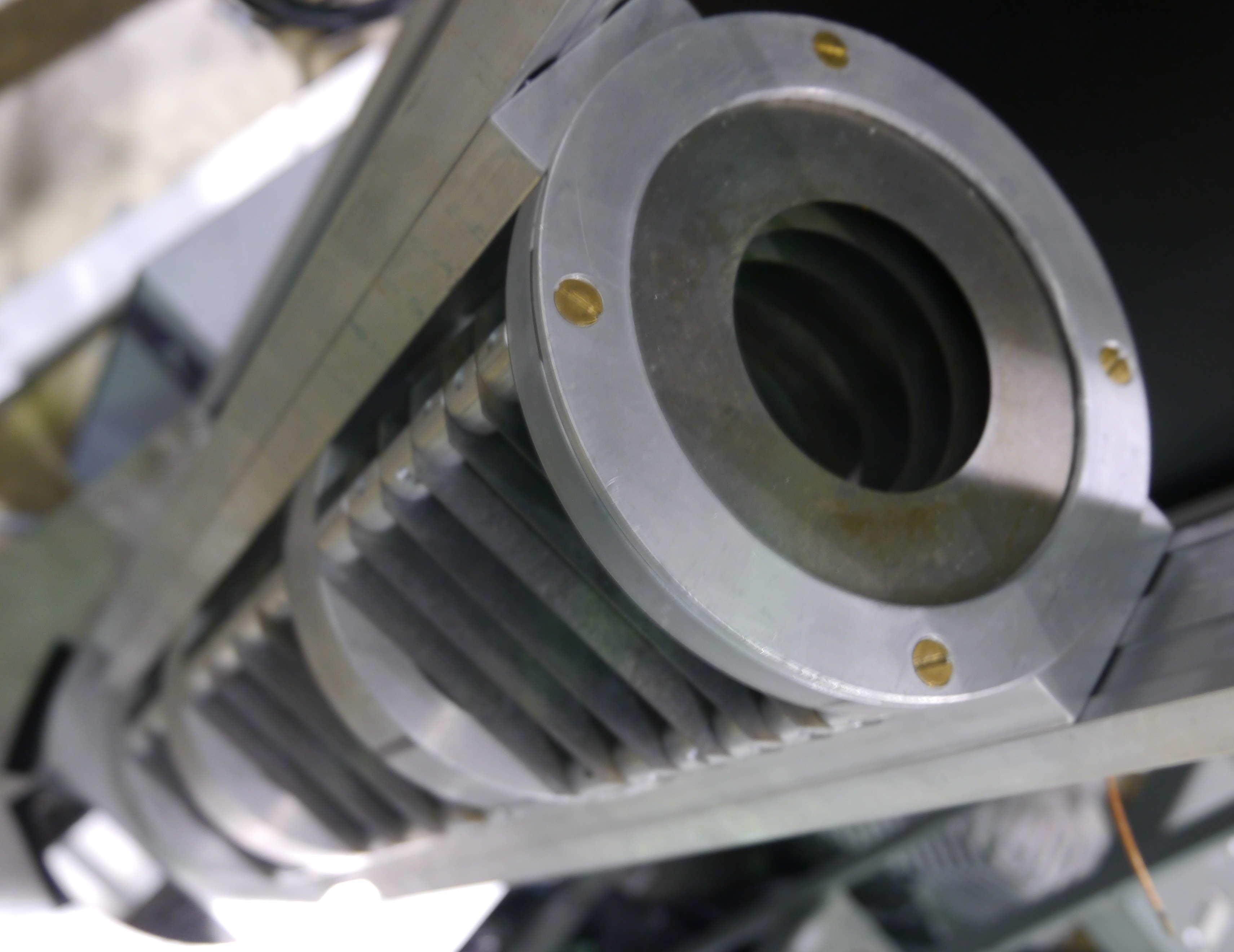}
\end{center}
\vspace{-0.25in}
\caption{\label{F:ECollimator} A view of the top of the electron collimator assembly.}
\end{figure}

The resultant electron collimator (Figure \ref{F:ECollimator}) consists of a series of seventeen 0.5 mm thick tungsten discs with 5.5 cm diameter circular apertures. The plates are supported in a holder that centers the apertures on the axis and spaces the plates in a non-periodic pattern designed to minimize the chance that a scattered electron can reach the detector.  The axial alignment of the apertures was checked optically using a theodolite and a set of reticles.  The centers of the measured apertures were concentric to within 0.2 mm. The electron collimator axis was aligned to within 1 mrad of the magnetic field axis.  The electron collimator assembly is mechanically fixed to the insert, which in turn is aligned to the axis of the experiment. It is worth noting that a misalignment of the electron collimator may affect the geometric acceptance function $f_a(E)$, but will not cause a false asymmetry.

\subsection{Proton Collimator}
The alignment of the proton collimator is more demanding than that of the electron collimator because a misalignment will directly cause a false asymmetry.  For any given electron momentum, the acceptance must be identical for the two coincidence proton groups (see Figure 2).  We require an alignment to within 0.1 mrad of the magnetic field axis. The axial momenta of the two groups are very similar due to the acceleration potential of the electrostatic mirror.  All protons make between 1 and 1.3 cyclotron orbits as they travel through the collimator.  The 8.00 cm diameter collimator extends from 39.5 cm to 179.5 cm above the center of the neutron beam.  It is machined from a monolithic aluminum tube with 55 apertures, spaced 2.54 cm apart, cut into the inside.  Each aperture has a knife edge shape to minimize scattering effects.

The proton collimator is rigidly fixed to the insert support structure.  Alignment is accomplished using optical reticles in the top and bottom apertures of the collimator structure.  

\subsection{Electrostatic Mirror}
\label{sec:mirror}
The electrostatic mirror surrounds the neutron decay region and provides a highly uniform vertical electric field that serves two purposes. First, it reverses the axial momentum of all protons whose initial trajectory points away from the proton detector.  Second, it increases the upward axial momentum of all protons toward the proton detector in order to reduce the difference in momentum between the two proton populations in the collimator. A transverse electric field can deflect the trajectories of the two proton groups differently and create a false asymmetry. Simulations showed that the transverse electric field in the mirror should be less than $10^{-3}$ of the axial field to reduce this effect to below \mbox{0.5 \%} of the $a$-coefficient. At the same time, the structure that provides the field must be relatively transparent to neutrons entering through the side and transparent to both electrons and protons exiting through the ends.

The boundary conditions for a right-cylindrical region containing a uniform electric field call for flat, parallel, equipotential ends with cylindrical side-walls that enforce a zero charge boundary condition. From an electrostatic point of view, the end boundary conditions could be met with parallel metal plates, but the end boundaries must also be transparent to protons. Accordingly, each end is constructed from a grid of fine parallel wires. In addition, the region above the mirror, all the way to the top of the proton collimator, must be free of electric field. Since the proton collimator and the surrounding vacuum can are at ground potential, the top grid must also be at ground. The wall boundary condition is somewhat harder to achieve. The potential on the walls must decrease linearly from the bottom of the mirror to the top.

\begin{figure}[htbp]
\begin{center}
\includegraphics[width = \bc_figure_width]{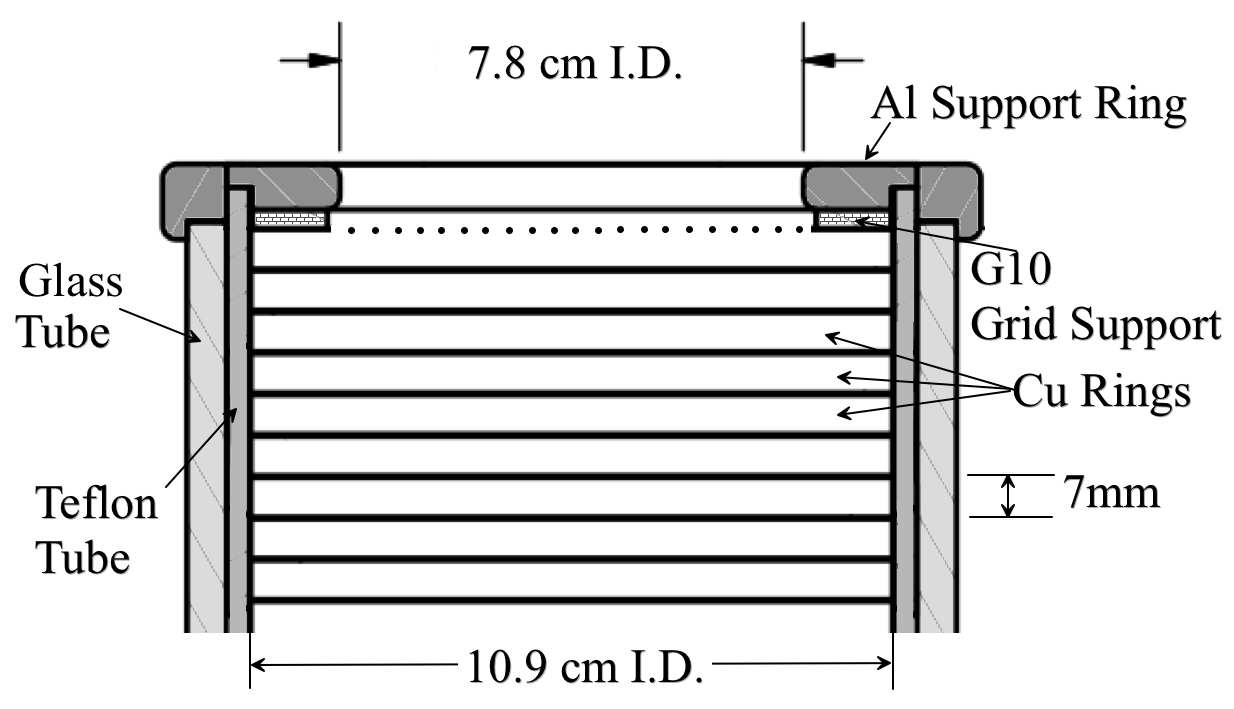}
\end{center}
\vspace{-0.25in}
\caption{\label{F:Mirror} Detail of top of mirror showing the printed circuit bands and the support structures. The grid wires are mounted on the bottom of the support, aligned with the middle of the gap between the top two Cu rings.}
\end{figure}

The electrostatic mirror wall consisted of a 0.25 mm thick sheet of Teflon, with a 4.5 $\mu$m thick layer of copper, divided into 63 parallel rings by photolithography \cite{Polyflon,DISCLM}. The rings have a periodicity of 7 mm including a 0.3 mm gap between rings (Figure \ref{F:Mirror}). The Teflon itself scatters about \mbox{1 \%} of the neutron beam and the copper plating scatters or absorbs about \mbox{0.1 \%} of the beam.

\begin{figure}[htbp]
\begin{center}
\includegraphics[width = \bc_figure_width]{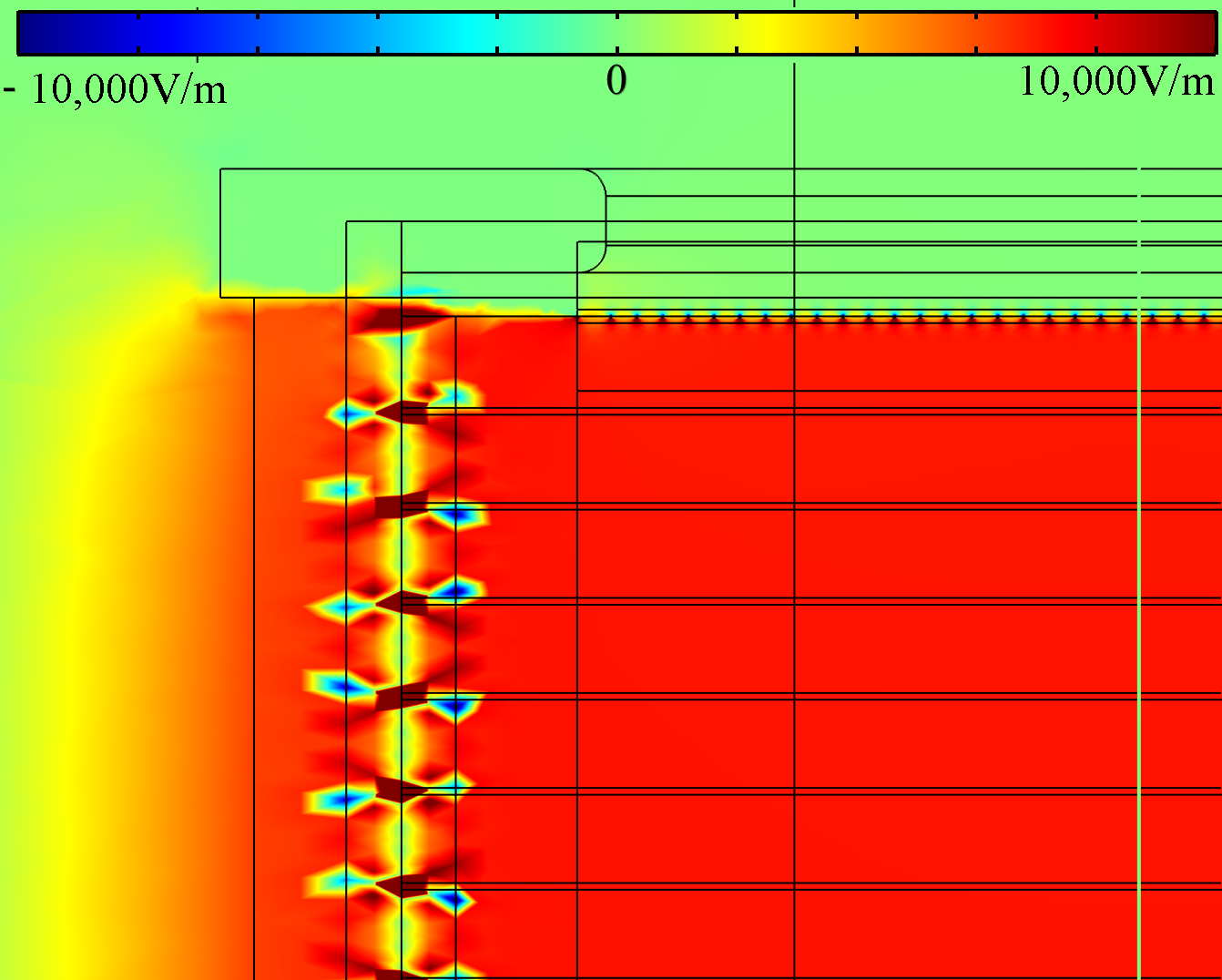}
\end{center}
\vspace{-0.25in}
\caption{\label{F:WallField} Finite element model of the magnitude of the electric field near the top of the electrostatic mirror. The strong non-uniformities caused by the 7 mm wide Cu bands and the gaps between them are seen to be confined to the region near the walls. Similarly, the strong fields produced by the grid wires are confined in a thin layer round the grid. }
\end{figure}

Each ring is held at a well-defined potential by a resistor chain attached to the outside of the rolled-up sheet. The resistors are chosen in a pattern of the form
$$ 500\;k\Omega, 1\;M\Omega,1\;M\Omega, \cdots 1\;M\Omega, 1\;M\Omega, 500\;k\Omega $$
The electric potential on the wall is thus a set of equal sized steps instead of the ideal linear gradient. These steps produce variations in the uniformity of the electric field, but these are confined to a region near the walls as can be seen in the calculated field map of Figure \ref{F:WallField}. The copper bands not only create the field inside the mirror but also shield the inner region from external fields. There are large electric fields just outside the mirror due to grounded supports that create fields up to 100 times larger than the field inside the mirror. The 0.3 mm gaps are small enough to ensure negligible penetration of the external fields (Figure \ref{F:WallField}). Monte Carlo simulations showed that these localized non-uniformities have no effect on the proton acceptance at the \mbox{0.01 \%} level. 

\begin{figure}[htbp]
\begin{center}
\includegraphics[width = 2.5in]{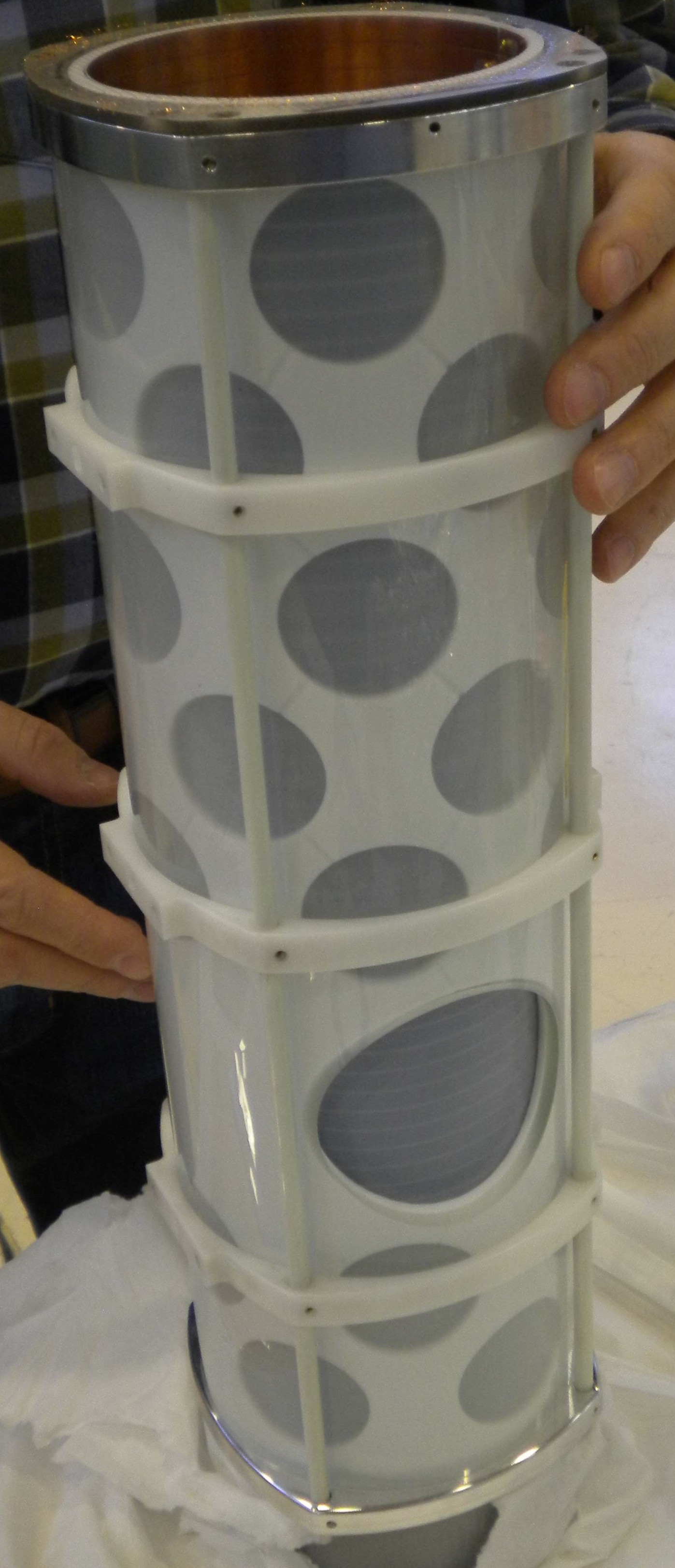}
\end{center}
\vspace{-0.25in}
\caption{\label{F:mirror} Electrostatic mirror assembly awaiting end grids. The Cu bands are visible through the thin Teflon sheet where it shows through the large holes in the thick Teflon tube. The large hole in both the thick Teflon and the glass support tube is visible between the lower pair of Teflon support rings.}
\end{figure}

Mechanical support for the copper-clad Teflon is provided by an 11.7 cm ID borosilicate glass tube, which also helps to absorb scattered neutrons, and by a Teflon tube with 4 mm thick walls. This Teflon tube provides reference surfaces at the top and bottom that mate with the grid supports, setting the grid position and separation to a tolerance of 0.1 mm. The thin Teflon slides into the thick Teflon tube and the whole assembly slides into the glass tube. The glass tube and thick Teflon tube have large holes to permit the neutron beam to pass in and out so that the beam passes only through the thin Teflon. A further array of approximately 2.5 cm diameter holes in the thick Teflon minimizes absorption of scattered neutrons in the thick Teflon (Figure \ref{F:mirror}). 

\begin{figure}[htbp]
\begin{center}
\includegraphics[width = \bc_figure_width]{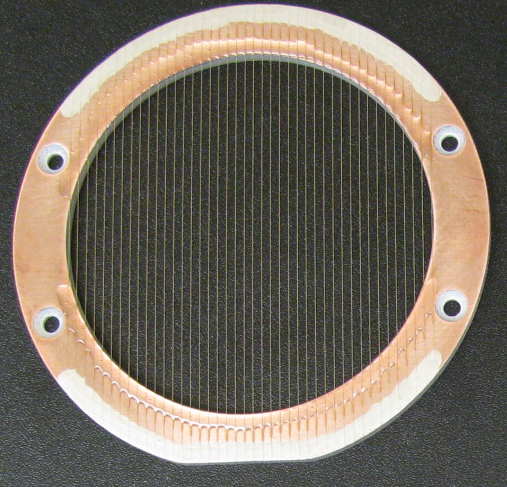}
\end{center}
\vspace{-0.25in}
\caption{\label{F:grid} End cap grid for the electrostatic mirror showing the 100 $\mu$m wires attached to the copper-clad printed-circuit board support.}
\end{figure}

The top and bottom ends of the cylinder are grids of 100 $\mu$m diameter wires spaced 2 mm apart on a support ring of copper-clad printed circuit board material (Figure \ref{F:grid}). The wires are attached to the rings with a thin (less than 0.3 mm) layer of epoxy cement and then the support rings are screwed to Al supports using flat-headed screws that lie flush with the copper surfaces, to avoid introducing irregularities in the electric field. The Al supports mate with the reference surfaces on the thick Teflon tube and position the plane of the grid wires at the level of the gap between the first and second copper bands on the mirror wall. Finite element models showed that this leads to the most uniform field. The two grids are held 42.7 cm apart by the Teflon tube.

The grid of wires allows approximately \mbox{95 \%} of the protons to pass through, but does not create the ideal boundary condition of an equipotential sheet. Several kinds of error are introduced. 

\begin{figure}[htbp]
\begin{center}
\includegraphics[width = \bc_figure_width]{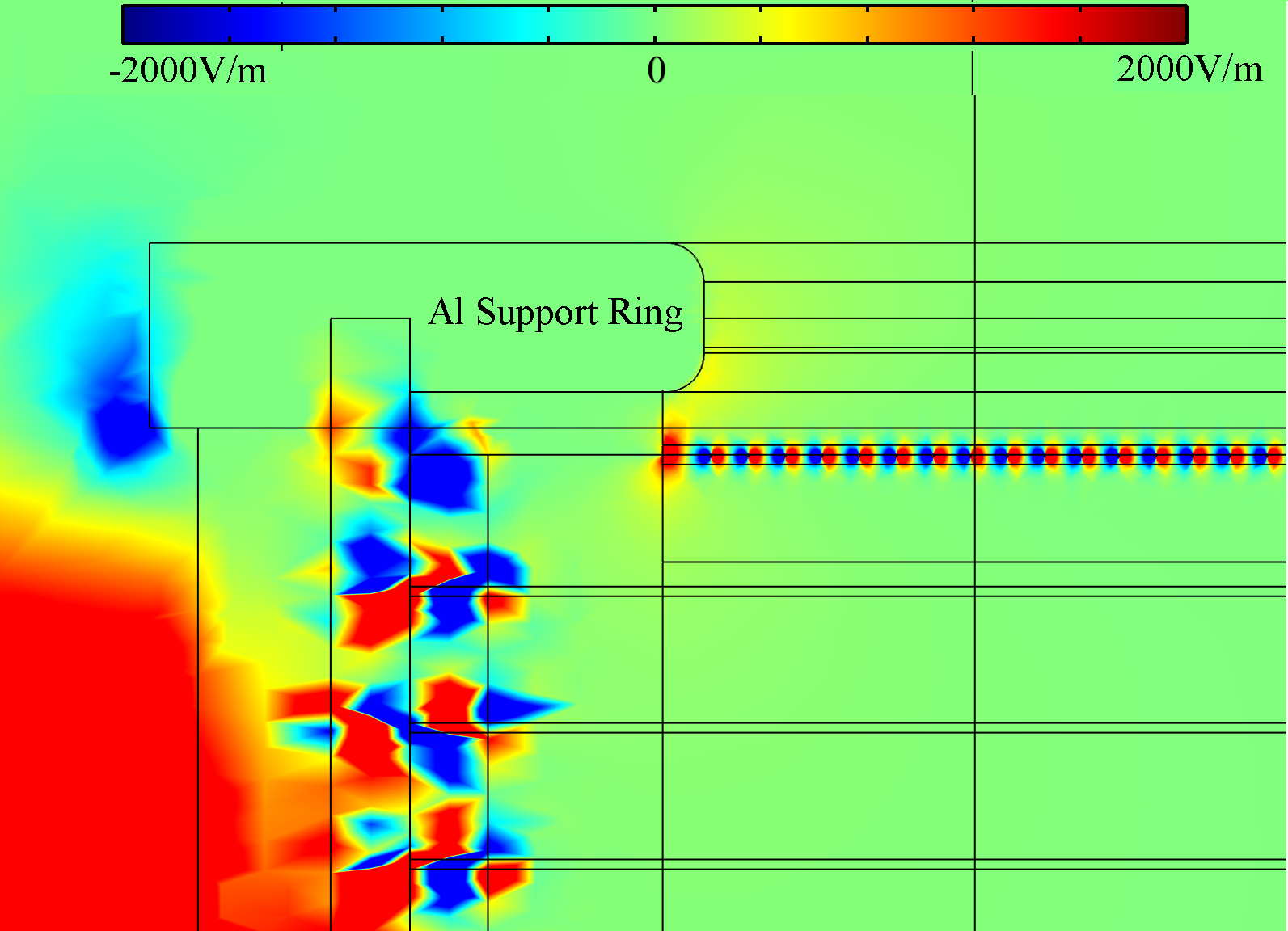}
\end{center}
\vspace{-0.25in}
\caption{\label{F:MirrorDetail} Finite element model of the transverse electric field in the region around the upper wire grid through which the protons exit the mirror. The strong small-scale fields around each wire are clearly visible. The shading around the curved edges of the Al support ring near the top of the figure is the most visible evidence for the large scale, low intensity, transverse fields that cause most of the false asymmetry.}
\end{figure}

\begin{itemize}
\item The small diameter wires produce strong transverse electric fields in the gaps between them through which the protons must pass. These fields are very strong but very localized.  In practice their influence tends to average out over the proton trajectories. 

\item The aperture in the grid support has a diameter of only 7.86 cm, slightly smaller than the 8 cm inner diameter of the proton collimator. The proximity of the grounded support to the grid wires also leads to moderate range transverse fields above the grid. Some evidence for these fields can be seen in Figure \ref{F:MirrorDetail}.

\item There are long range transverse fields both above and below the grid that are the result of the the gaps between the wires that make up the grid.
\end{itemize}

The long and medium range fields affect acceptance for the two proton populations in slightly different ways and lead to a systematic error in the proton asymmetry that will be discussed in section \ref{F:ESMirrorCorrect}.

\subsection{Insert Alignment}
The insert and its sub-assemblies are aligned to the experimental axis in a multi-step process. First the insert is placed on a test stand and Brunson 6150\cite{DISCLM} optical target reticles are placed in the top and bottom apertures of the proton collimator and in the top and bottom apertures of the electron collimator. The reticles are observed through a theodolite mounted on an x-y translation stage that views the insert axis through a pentaprism mounted on a vertical translation stage. Since the dimensions of the marks on the reticles and all the distances are known, angle measurements with the theodolite can be converted to distances. It is possible to measure all translations of the reticles to a precision of ~25 $\mu$m. The individual sub-assemblies are shimmed until the proton collimator, electrostatic mirror and electron collimator are aligned to the same axis to within 0.1 mrad. 

Once the insert components are aligned, the insert is mounted in the vacuum vessel. As described above, the magnetic field is aligned to the axis determined by a pair of precision alignment bearings attached to the main vacuum vessel around which the field mapper rotates. Since the lower bearing must be removed before the insert can 	be put in place, a second optical procedure was required to ensure that the insert is aligned with the respect to the magnet axis. First, reticles are installed and centered in the top and bottom precision alignment bearings. The positions of the reticles are determined with the theodolite and pentaprism and recorded. The reticles and the bottom bearing are then removed and the insert lowered into place, with reticles in the bottom aperture of the electron collimator and the top aperture of the proton collimator. The position of the (fixed) bottom reticle is measured and the position of the top reticle adjusted with the spider to minimize the misalignment of the insert axis with respect to the axis determined by the bearings. Measurements are typically made by at least two independent observers. On some occasions the insert was removed and replaced to verify that the bottom position was repeatable to within 50 $\mu$m. The whole insert was aligned to within 130 $\mu$m, corresponding to an angular precision of 0.09 mrad.

\section{\label{PDet}Proton Detector}
After pre-acceleration in the electrostatic mirror and collimation in the proton collimator, protons are accelerated into a 600 mm$^2$, 1000 $\mu$m thick, surface barrier detector mounted $\approx4$ cm to one side of the beam axis.  If the proton detector were on the experimental axis, decay electrons could scatter from the detector and impart a signal in the electron detector at the opposite end of the apparatus. This would result in a signal that would confound the ability to extract the correct wishbone asymmetry.  To avoid this, the proton detector is placed off axis, with the protons guided and focused by electric fields.  Most of the electrons, with far more rigid trajectories, pass by the detector onto a polyethylene plate in the low field region, leaving them unlikely to find their way back through the proton collimators and into the beta spectrometer.  Due to the acceleration in the electrostatic mirror, proton trajectories above the collimator are approximately independent of the wishbone branch of the decay, but there remains a slight dependence so it is important for the proton detector efficiency to be as close to 100 \% as possible. The final electrode configuration was reached by simulating proton trajectories through computed electric fields using the AMaze\cite{DISCLM,AMAZE} finite-element package  to produce a system in which all simulated protons reached the active surface of the detector. 

\begin{figure}[htbp]
\begin{center}
\includegraphics[width = \bc_figure_width]{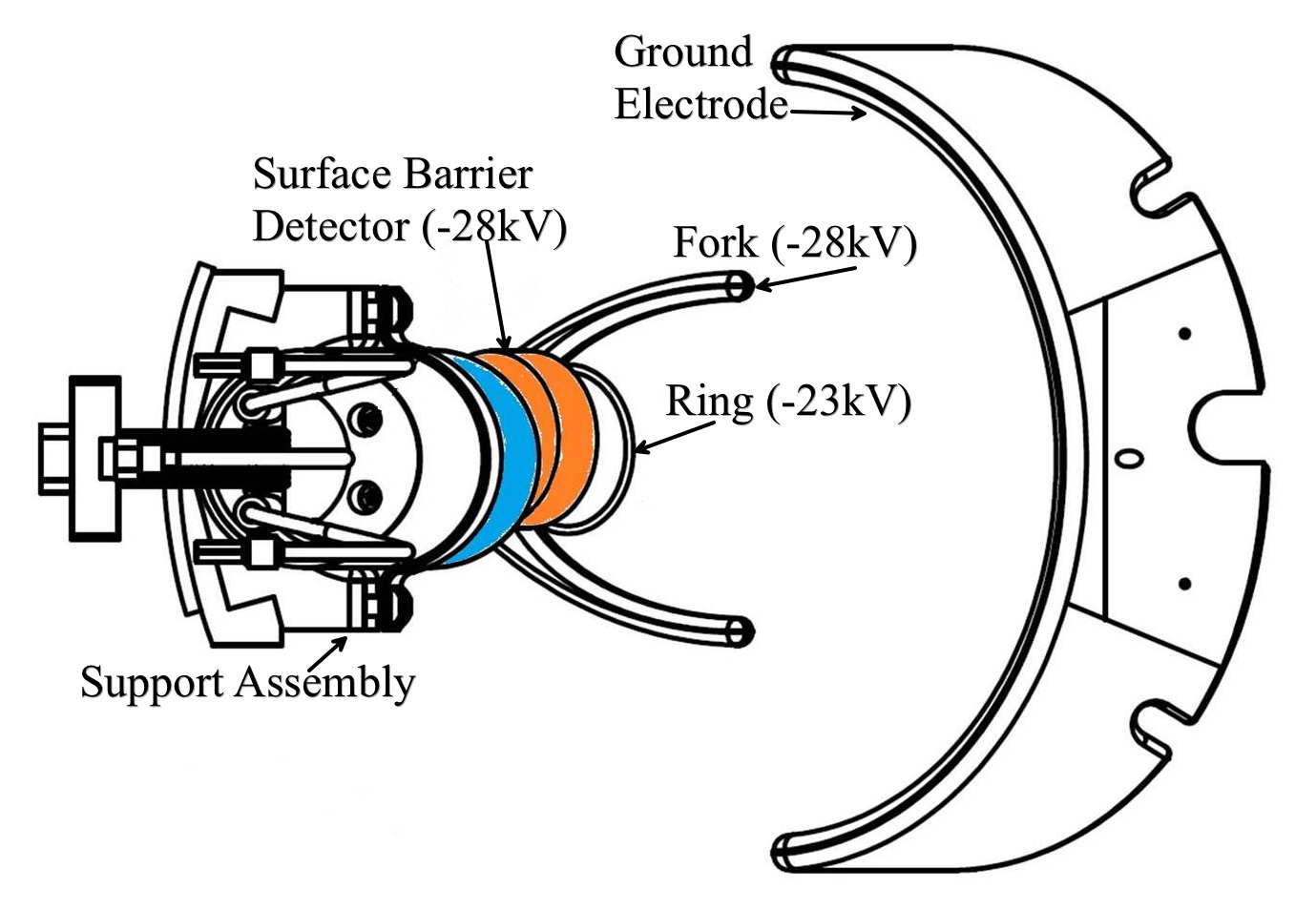}
\end{center}
\vspace{-0.25in}
\caption{\label{F:PDetAbove} Schematic view of proton detector from above, showing the three field shaping electrodes and their relation to the surface barrier detector.}
\end{figure}

\begin{figure}[htbp]
\begin{center}
\includegraphics[width = \bc_figure_width]{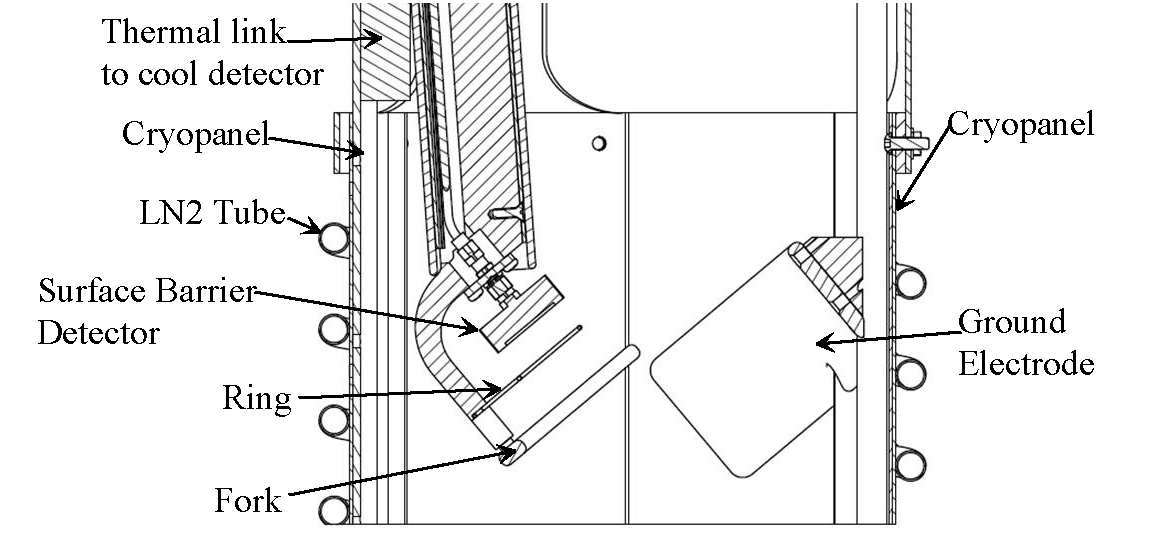}
\end{center}
\vspace{-0.25in}
\caption{\label{F:ProtDetSide} Schematic view of proton detector from the side, showing the spatial relationships of the field shaping electrodes and the cryopanels that support and cool the detector. The seam in the center of the picture shows the position of the proton collimator axis.}
\end{figure}

Figures \ref{F:PDetAbove} and \ref{F:ProtDetSide} show the resulting proton detector assembly. In order to insure that protons travel through the dead layer, the surface barrier detector is held at -28 kV near two electropolished focusing electrodes. Just in front of the detector there is an aluminum ring held at -23 kV. It has a diameter of 31.5 mm and is formed from 1.5 mm diameter wire. Slightly further away lies the fork, a wishbone-shaped electrode held at -28 kV. The tines are 6.3 mm thick, 74.2 mm long, and are separated at their tips by 65.3 mm.  These electrodes, together with the grounded cryopanels that surround the region and the large curved ground electrode attached to them, steer the protons to the detector while letting the electrons pass through.

An inner Cu cylinder held at -28 kV containing the high voltage wires leads from the top of the vacuum chamber to the detector and electrodes, as well as to a spark protection circuit. It is inside a Teflon cylinder that acts as an electrostatic insulator and thermal link.  The detector is cooled via thermal contact with the cryopanels to minimize the thermally induced leakage current.  Because Teflon is also a thermal insulator, an aluminum clamp thermally coupled to the cryopanels is used to optimize thermal conduction as much as possible.  In this way, temperatures as low as 206 K are achieved at the detector.  The wires from the detector lead to a feedthrough flange and are connected to a preamplifier that sits outside the vacuum chamber at room temperature. All of these components are at -28 kV.

\begin{figure}[htbp]
\begin{center}
\includegraphics[width = \bc_figure_width]{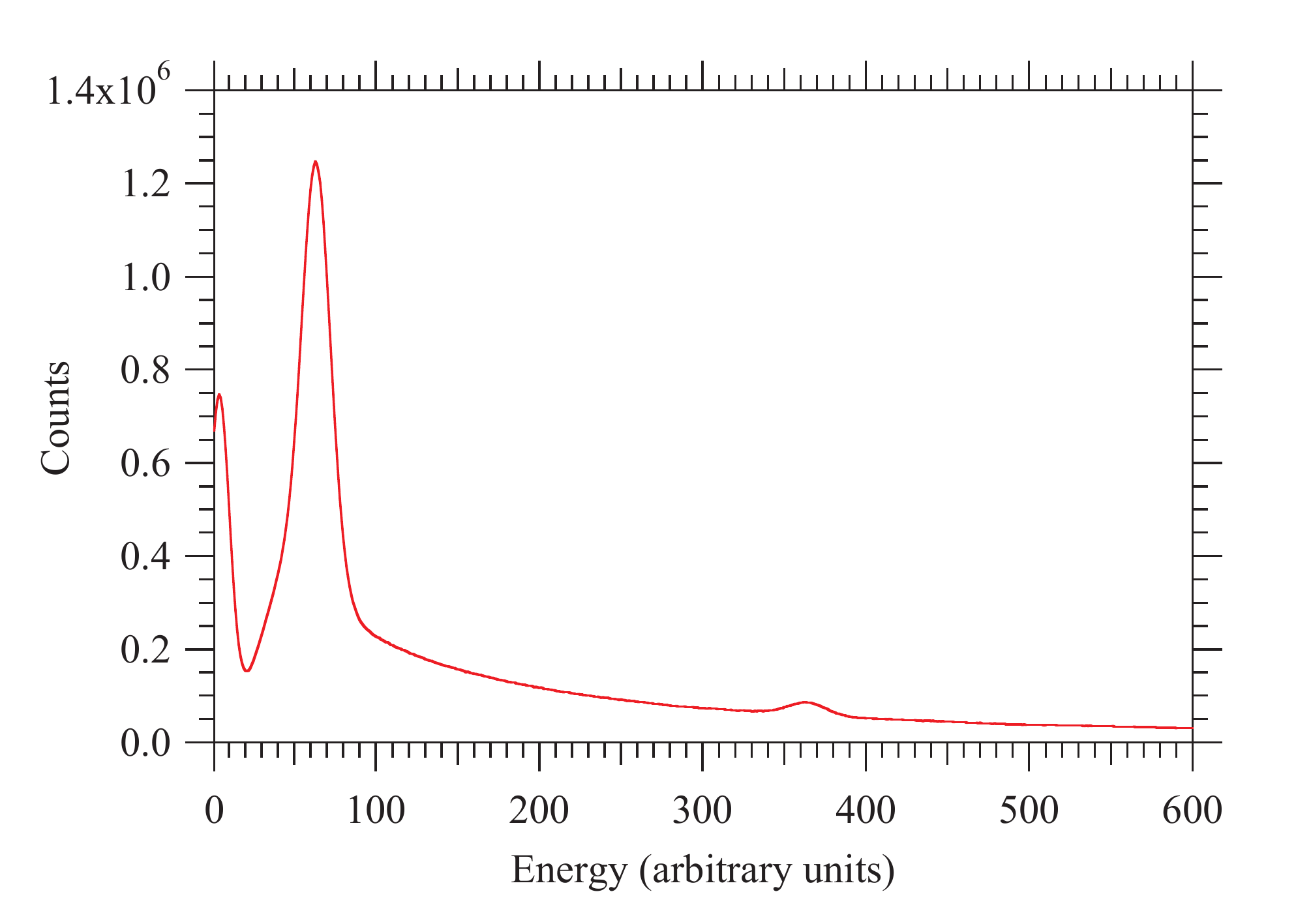}
\end{center}
\vspace{-0.25in}
\caption{\label{F:PSpectrum} aCORN proton spectrum. The large peak lying between channels 30 and 90 contains the decay protons. The small peak near channel 360 is from a pulser built into the preamplifier.}
\end{figure}

The preamplifier was designed with the goals of (1) having the preamplifier operate at room temperature while connected to the detector through about 35 cm of low-capacitance coaxial cable, (2) choosing a design engineered to reduce noise and assembled with low noise components, (3) protecting the preamplifier from high voltage sparks through a combination of shorting capacitors and gas filled high voltage shorts, and (4) running an on-board pulser at a rate of 1 Hz with an amplitude well above the proton signal. The resolution of the silicon detector with the new preamplifier was measured using 59.5 keV gamma-rays from the decay of $^{241}$Am and found to be 1.27 keV. These preamplifiers operated successfully for aCORN with no failures for three years.  Since the preamplifier operates at the -28 kV detector potential, the output signal is converted into a light pulse, which travels via optical fiber to the grounded data acquisition system.  Figure \ref{F:PSpectrum} shows a raw proton spectrum.

Both ring and fork are adjustable.  Alignment was accomplished using an alignment jig with a mirror polished surface on the axis of symmetry.  Matching the reflection to the parts of the fork that were visible insured that the fork was aligned.  

\begin{figure}[htbp]
\begin{center}
\includegraphics[width = \bc_figure_width]{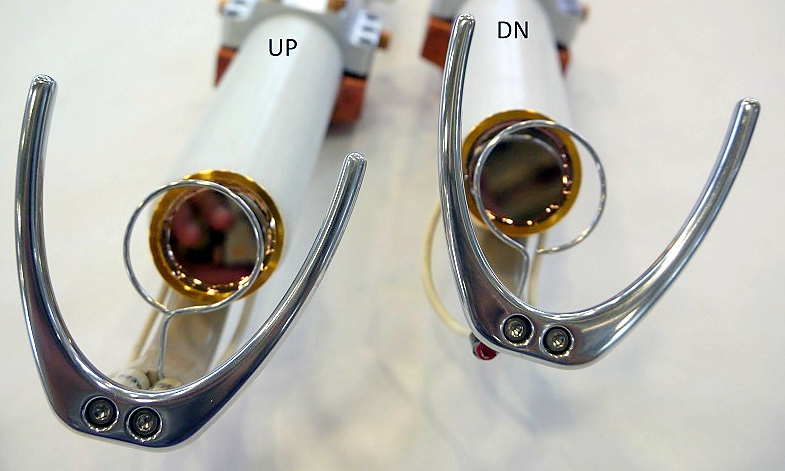}
\end{center}
\vspace{-0.25in}
\caption{\label{F:PDetPhoto} Photograph of the two aligned detector assemblies, one for the field up configuration and one for the field down, showing the slight reorientation of the fork and ring relative to the detector.}
\end{figure}

To address the possibility of the neutron beam having a slight net polarization, which would cause a systematic error in $a$, data were taken using both orientations of the magnetic field.  For each magnetic field direction the proton trajectories are displaced slightly in the azimuthal direction due to the $\vec{B}\times\vec{E}$ effect. This effect was accommodated by a small shift of the detector and ring electrode whenever the magnetic field was reversed.  Figure \ref{F:PDetPhoto} shows the two different aligned detector assemblies.

\section{\label{betadet} Backscatter Suppressed Beta Spectrometer}
\par
The beta spectrometer was used to measure both the detection time and energy of the electrons arising from the neutron beta decay. We require a detector that is able to detect with high efficiency electrons whose momenta were within the acceptance determined by the axial magnetic field and electron collimator. The electron energy response should be linear and measured with a calibration uncertainty of less than \mbox{2 \%}. Electrons that backscatter from the detector without depositing their full energy cause a particular systematic problem in aCORN (see section \ref{systematics}). The spectrometer was designed to suppress such events.
\par
\begin{figure}[ht]
    \centering
    \includegraphics[width=\bc_figure_width]{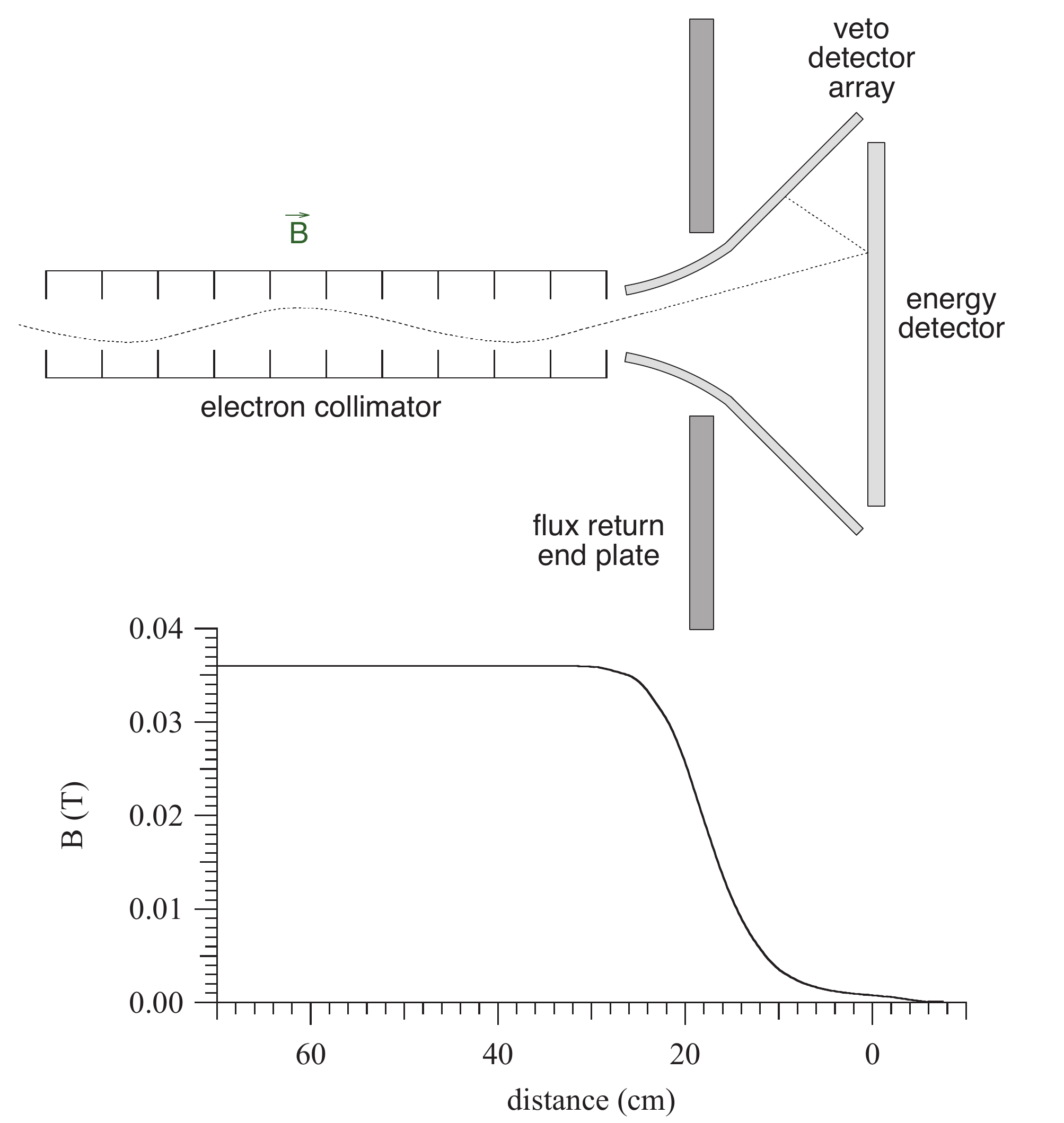}
    \caption{Top: design scheme of the backscatter suppressed beta spectrometer. Beta electrons pass through the beta collimator in a uniform
    axial magnetic field, into the spectrometer and onto the energy detector. Electrons that backscatter are likely to strike the veto detector array because the 
    magnetic field is much weaker at the position of the energy detector.  Bottom: The aCORN axial magnetic field aligned to the top plot.}
    \label{F:bsscheme}
\end{figure}

The design principles of the beta spectrometer are illustrated in Figure \ref{F:bsscheme}. The spectrometer was mounted at the end of the electron collimator, below the iron flux return plate. Electrons that passed through the collimator were transported by the magnetic field and efficiently admitted into the beta spectrometer via the opening in the veto array. All such electrons with kinetic energy $>$100 keV struck the active energy detector, a circular sheet of plastic scintillator, 0.5 cm thick. The magnetic field was significantly reduced at the position of the energy detector, so electrons that backscattered from it were unlikely to return through the entrance without striking the veto detector, an octagonal array of eight plastic scintillator paddles. The backscatter rejection efficiency was designed to be approximately \mbox{90 \%}. This was confirmed by estimates from neutron decay spectra.

\par
\begin{figure}[ht]
    \centering
    \includegraphics[width=\bc_figure_width]{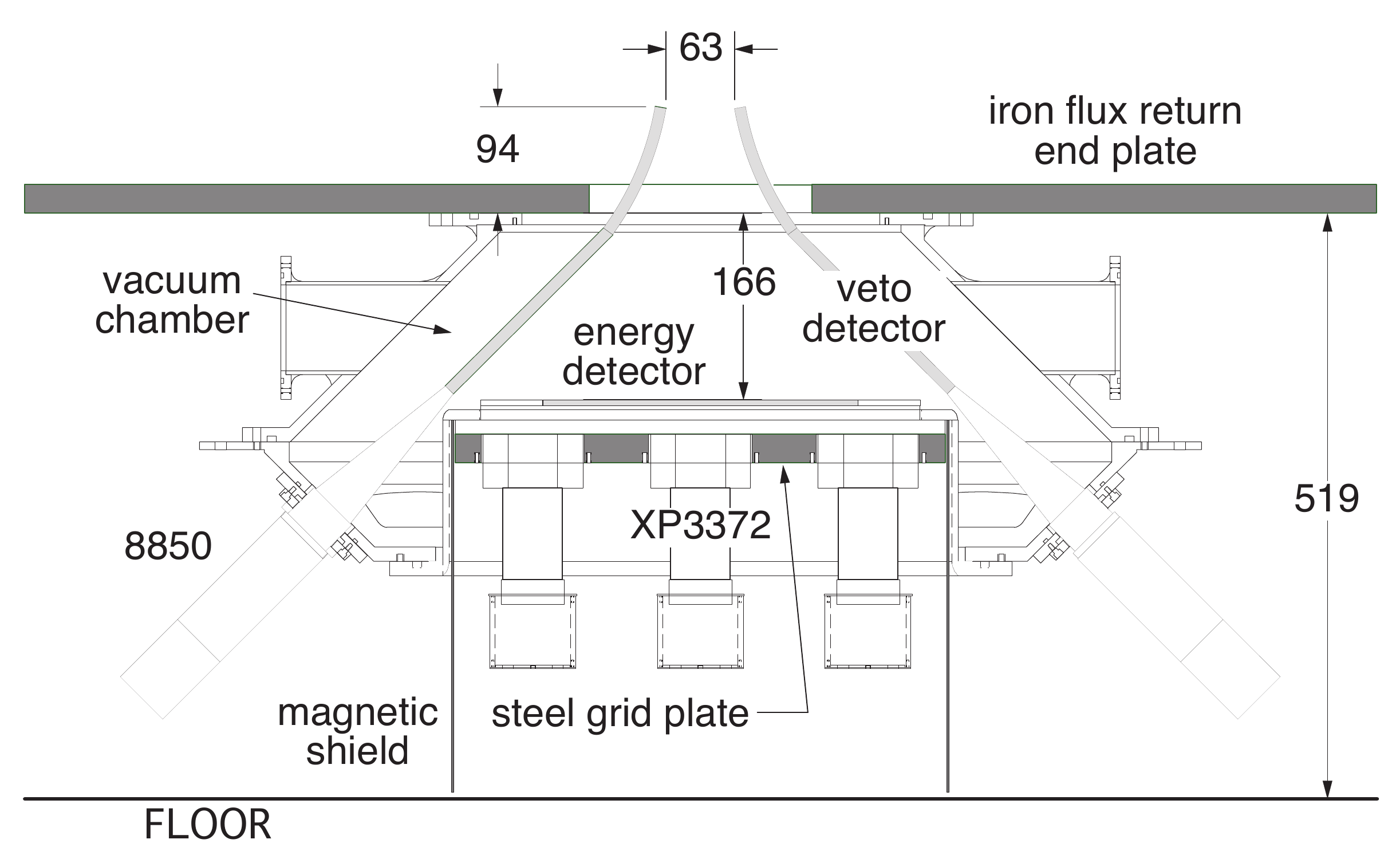}
    \caption{General arrangement of the spectrometer showing the major components (dimensions in mm).}
    \label{F:bsCutaway}
\end{figure}

The physical arrangement of the spectrometer is shown in Figure \ref{F:bsCutaway}. The energy detector was backed by an acrylic light guide that transported scintillation light to an array of 19 7.6 cm (3-inch) hexagonal, 8-stage photomultipliers \cite{PMT} (PMTs) arranged in a honeycomb pattern. The energy light guide, supported by a steel grid plate that positions the PMTs, also functioned as the vacuum window. The veto detector was a closed octagonal array of eight veto paddles, each consisting of two trapezoidal sections, one flat and one curved, glued to an acrylic light guide that penetrated the vacuum chamber. Each veto light guide was coupled to a 5.1 cm (2-inch) circular 12-stage PMT \cite{VetoPMT}.  The scintillator and light guide of each veto paddle was wrapped in a single layer of thin aluminized mylar for optical decoupling from the energy detector. High voltage for all 27 PMTs was supplied by a computer-controlled 32 channel Wiener ISEG\cite{DISCLM} high voltage system.

\par
A more detailed description of the aCORN beta spectrometer, measurements with conversion electron and neutron decay sources, energy calibration results, and the backscatter rejection efficiency, can be found in reference \onlinecite{bsNIM}.

\section{\label{DAQ}Data Acquisition}
The aCORN apparatus generates pulses from the 19 energy measuring PMTs, 8 veto PMTs, and
the single, surface-barrier proton detector. All of these signals feed into two PIXIE-16 \cite{PIXIE,DISCLM} modules. These modules digitize the
signals using 12-bit 100 megasample per second analog to digital converters (ADCs) and extract pulse time and pulse
height information by digital filtering of the signals. Each pulse
recognized by the system produces a 16-byte record including the time,
to a precision of 10 ns, and a 15-bit pulse height. The system can handle up
to about 7 events per 10 ns time-slice. A single electron hitting the
scintillator may produce more than 19 events, which the system then spreads out
over a number of 10 ns time-slices. 

The PIXIE system imposes an unusual dead-time structure on the electron detection. First, entire modules shut down when their memories are full and data must be transferred to the host computer, and second, each individual channel has a dead time of between 200 ns and 300 ns during which it cannot accept another pulse. The first of these effects simply reduces the data rate, so long as both modules shut down together, however the second effect is more serious. Because the aCORN beta detector uses 19 PMTs to measure the electron energy, many events illuminate only some fraction of the PMTs. It is possible to have an electron event in which some PMT events are recorded correctly but others are missed, because the PIXIE channel is still in recovery. This results in erroneously small energies for those electron events, exactly the problem that the veto-suppressed electron detector was designed to avoid. This problem is corrected in the analysis phase by imposing a 500 ns uniform dead-time after each electron event. By contrast, the proton detection is much simpler. There is a single proton energy pulse to be measured in each module. Because this pulse is of longer duration than the PMT pulses, the processing is slower and results in a simple 3 $\mu$s dead-time for the proton signal.

The PIXIE firmware was modified by XIA\cite{DISCLM} to record an event only if it were seen by at least two of the channels within a 100 ns window. Since any electron event in our usable energy range will be detected by at least two PMTs, this effectively suppressed noise and dark current from individual PMTs. The proton signal was duplicated and fed into two PIXIE inputs on different PIXIE modules, so that proton signals were acquired only when both modules were active.  

The PIXIE creates a rather imprecise energy threshold on individual channels. This made it difficult to distinguish between decay protons and the low-energy noise. For some of the later data runs a hardware discriminator was added to the proton signal and both the raw energy signal and the discriminated signal recorded. Such data runs produced four separate events from each proton, an energy signal from each module and a discriminator signal from each module.

The output from the PIXIE is a list of individual channel events consisting of a channel identifier, a 48-bit time, and a 15-bit energy. In normal operation the experiment produces approximately 1 Terabyte of raw data per 24 hours. For the first half of the experimental runs data were collected for several minutes, filling the computer memory, and then they were dumped to disk for later reduction and analysis. This wasted about \mbox{15 \%} of the live beam time and created a serious data storage problem. For the latter half of the runs, the acquisition software was rewritten to interleave data collection, disk writing, and the first stage of the analysis, resulting in a 20-fold decrease in the amount of data that must be archived and in the elimination of the \mbox{15 \%} time loss. For quality assurance purposes, we continued to record occasional raw data files.

In addition to the high-frequency event data, a large number of
experimental parameters were recorded at much lower frequency. These
included the PIXIE parameters, the currents to the various magnet
coils, the proton-detector power-supply voltages, sampled values of
the magnetic field, and the temperatures of the cryopanels. These were
all written to disk at two minute intervals and archived with the
PIXIE data.

\section{\label{reduction}Data Reduction}
The data from the PIXIE consist of individual pulse energies from the
19 electron PMTs, the 8 veto PMTs, two copies of the proton detector energy, and two copies of the energy-discriminated proton signal. The energies of all the electron PMTs are summed to yield a total electron energy for the event and all of the data are written to disk, along with a header that contains the running parameters for the data set.
A data
bottleneck within the PIXIE means that the events are not strictly
time-ordered, but each event carries an accurate time stamp that we use
to restore time order. Most individual PIXIE pulses arise not from
neutron decay signals but from background, mostly secondary events produced by neutron capture prompt gamma rays.

We normally store only those events which lie within a small window
around a proton event, a process that we call distilling. We chose to
collect data for 10 $\mu$s before each proton event, somewhat longer than the maximum proton flight time, and 1 $\mu$s after, when there can be no decay-related electrons. In this way, we can study the background spectrum all around the proton event.  Since there are no protons in or near the data that are discarded, the discarded data can consist only of background events.

The distilled data contain raw detector events, each
consisting of an energy and a time stamp. They must still be assembled
into candidate decays. This is done in a second, off-line, analysis
phase, which we call reducing. The reducer searches through the
time-ordered events from the distiller and builds decays in a
three-step process. First, the events are separated into PMT events,
proton events, and discriminator events. Second, the PMT events are
assembled into complete electron events or discarded as noise, and
the proton and discriminator events are grouped into complete proton
events. Third, each complete electron is associated with every proton
event that arrived with a 10 $\mu$s window after the electron and each complete event written to disk in a human-readable text format.

Because all the data must pass through the distiller and reducer, and because we were very concerned with the integrity of the data, the programs were
subjected to extensive testing. Two different versions of each 
program were written from scratch, using
different algorithms and different programming languages. The
two independent versions were then fed hundreds of gigabytes of data and
their outputs compared. The two independent systems returned identical 
results, giving us confidence in their integrity.

\subsection{Electron Events}
A single electron may generate anywhere from two PMT events up to 19
electron PMTs and 8 veto PMTs. Delays within the DAQ system spread the
individual PMT events over a time interval that is small compared to
the average time between electrons. The assembly process
starts with a single electron or veto PMT event. The reducer collects
all succeeding electron or veto PMT events within a 120 ns window.
It then scans ahead for a further 50 ns and rejects any electron group which has a PMT event within that dead time. This avoids misidentifying two closely-separated electrons as a single higher energy electron.

\subsection{Proton Events}
A single proton detection may result in up to four events in the data stream, a
proton energy signal from each of the two PIXIE modules, and a discriminator
signal from each PIXIE module. Like the electron events, these are grouped by
time. Since the discriminator pulse is delayed by approximately 1 $\mu$s from the associated energy signal, we require that
the proton events from the two PIXIE modules arrive within 100 ns of
each other and that the proton and discriminator events arrive within
1.5 $\mu$s. A proton with an energy well above the PIXIE threshold normally produces all four signals, while a proton that is accepted by the hardware discriminator but rejected by the PIXIE because of its soft threshold will produce only the two discriminator events.
The proton rate is low enough that event pileup is negligible.

\subsection{Electron-Proton Coincidences}
Because most electron events arise from the background and not from neutron decays, it is impossible to uniquely associate each electron with the correct proton. Instead, we
associate each proton with every electron that arrives within 10 $\mu$s prior to the proton. This means that individual protons and electrons may appear multiple times in the output, if there
are closely spaced protons. A typical proton is associated with
between 1 and 5 electrons. At most one of those electrons is the
result of a neutron decay (the decay rate is slow enough that the
probability of two decays within the 10 $\mu$s window is ${\approx 10^{-5}}$). The
remaining electron groups are due to electronic noise and background and are randomly
distributed. Except for the calculable effects of PIXIE dead-time, this results in a background level in the wishbone plot that does not depend on electron-proton time-of-flight.

\section{\label{systematics}Systematic Considerations}
\par
We are aware of a number of systematic effects that could influence the data. This section will describe these and discuss our strategies to measure and correct for them.

\subsection{Electrostatic Mirror}
Ideally, the electric field would be perfectly uniform inside the mirror and exactly zero outside it. Several effects lead to deviations from this ideal. First the end surface boundary conditions are provided by grids of parallel wires, all running in one direction. These introduce several imperfections in the field. The gaps between wires allow some field lines to pass through the end surface before reaching ground, reducing the axial field inside the mirror and increasing it outside. In an alternative, and equivalent, view, the small but finite size of the wires leads to very strong fields in their near vicinity that must be compensated by reduced fields further away in order that the total potential drop along the lines be correct. In either view, the result is transverse field components caused by the bends in the field lines. These take two forms: very localized, very strong fields close to the wires, and much longer range, weaker fields further from the wires, see Figure \ref{F:MirrorDetail}. The effects of the localized fields tend to average to zero along the length of a proton track but the longer range fields can affect the two proton populations to different extents and lead to a false asymmetry. This effect is made worse by the small diameter of the top grid, which exposes detectable protons to the strongest transverse fields that are found near the edges of the aperture. Deviations from the ideal electric field caused by the finite potential steps at the wall do not extend significantly in to the region of proton transport so this is not a systematic concern.

\begin{figure}
\begin{center}
\includegraphics[width = \bc_figure_width]{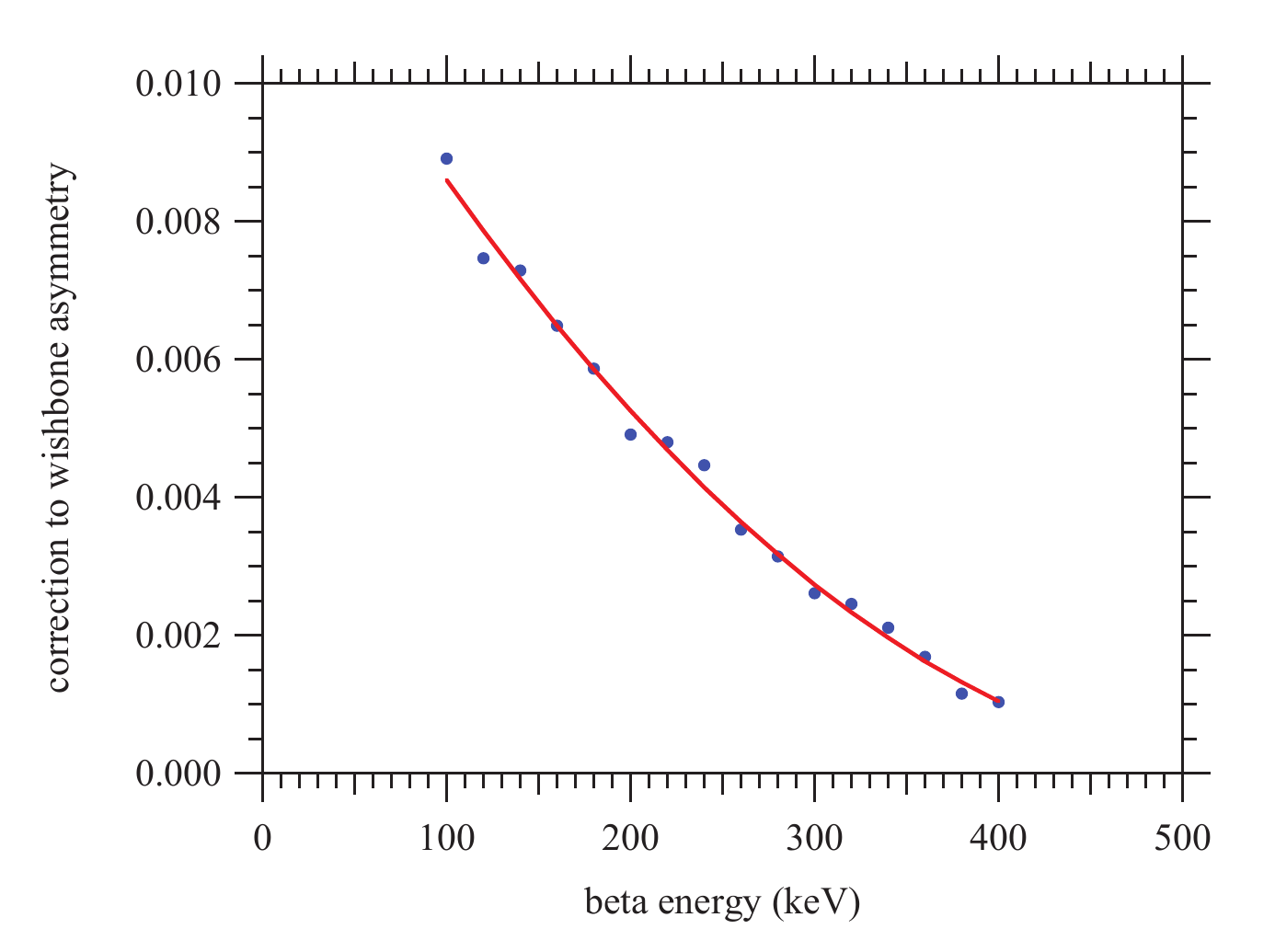}
\end{center}
\caption{\label{F:ESMirrorCorrect} The Monte Carlo calculated correction to the wishbone asymmetry due to the modeled transverse
electric field in the electrostatic mirror. The smooth solid line is the best fit to a second order polynomial.}
\end{figure}

We rely on detailed computer models of the electrode system and its fields to correct the data for the actual field. Accordingly, a highly detailed 3D finite-element model of the electrode system was built in COMSOL\cite{COMSOL,DISCLM}, including the finite widths of the wall bands, the finite diameter of the grid wires, and details of the grid support geometry. A 3-D electric field map generated from the COMSOL model was input into a Monte Carlo simulation of proton transport in aCORN to calculate
the effect on the wishbone asymmetry. Figure \ref{F:ESMirrorCorrect} shows the results, a correction that must be added to the wishbone asymmetry at each beta energy. 
The average size of this correction is \mbox{5.4 \%}.

\subsection{Proton Soft Energy Threshold}
As mentioned in section \ref{DAQ}, the PIXIE does not create a well defined threshold energy for proton pulses. Rather than a sharp threshold function, the threshold is ``soft'', acting over a small range of proton energy. Since this will preferentially affect lower energy protons, it will lead to a greater reduction in the slow proton population and thus possibly to a false asymmetry. Part way through the experimental run, a hardware discriminator was added to the system to obviate this problem. From that point all data included both the proton energy signal and the discriminator signal, which could be used to apply a rigorous energy cut in the analysis.

\begin{figure}
\begin{center}
\includegraphics[width = \bc_figure_width]{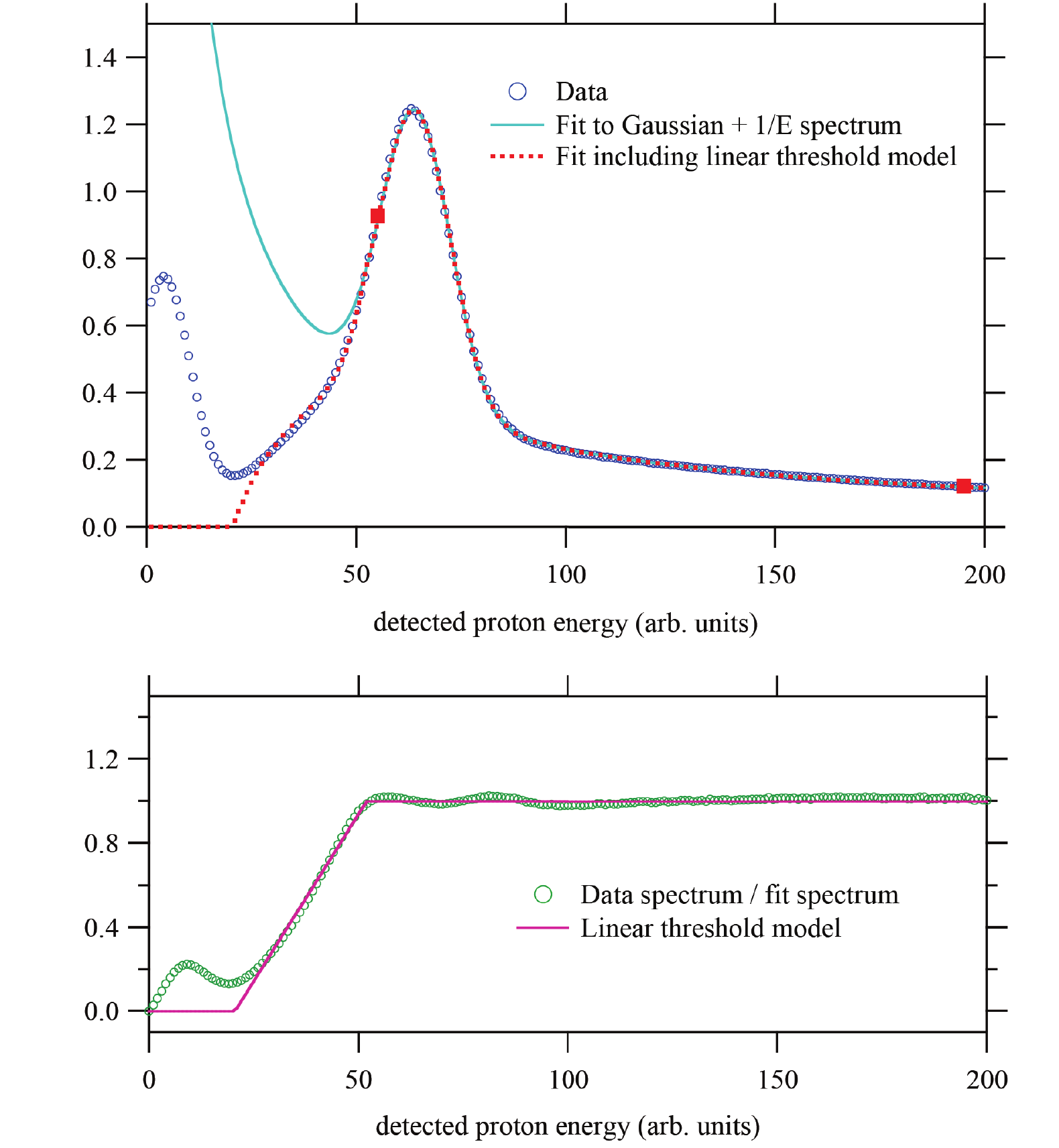}
\end{center}
\caption{\label{F:pSoftCalc} Top: Proton energy spectrum data, fit to a Gaussian peak plus 1/Energy background function in the region between the square markers, well above
the threshold region. Bottom: Data divided by the fit spectrum. In the threshold region (below energy channel 50) a linear threshold effect can be seen. The solid line is a simple linear model of this threshold effect. When this linear model is included in the fit function, the dashed curve in the top plot is obtained.}
\end{figure}

A model of the soft threshold effect was obtained using the following procedure. As shown in Figure \ref{F:pSoftCalc}, the measured proton energy spectrum was fit to a Gaussian peak plus a 1/Energy background function in the energy region well above the threshold. The peak energy here is about 29 keV due to acceleration of protons in the electrostatic mirror and the negative potential of the proton detector. The full energy spectrum, including the threshold region, was then divided by the fit function. This gives a measure of the PIXIE threshold function, which is approximately linear as seen in Figure \ref{F:pSoftCalc} (bottom). A simplified model of this threshold function was then included in the Monte Carlo simulation of the experiment in order to estimate the relative effect on the wishbone asymmetry, which was found to be \mbox{2.9 \%}.
 
\subsection{Absolute Magnetic Field Calibration}
The absolute value of the magnetic field affects our measurement through the computation of the $\phi$ functions of equation \ref{E:phis}. The main magnet current was monitored with a shunt resistor and showed variations of \mbox{$\pm0.5 \%$}, corresponding to variations of $\pm0.18$ mT. The absolute field was measured on several occasions by $^3$He NMR and variations at that level were observed. Monte Carlo simulations were made over the range of observed absolute fields and used to compute the resulting contribution to the systematic uncertainty.

\subsection{Magnetic Field Shape}
Because we had to reduce the magnetic field in the region of the proton detector in order to make the system stable, the magnetic field is not constant through the whole proton collimator. The actual shape of the magnetic field was computed using a 3-D finite element model in COMSOL, including the effects of the iron yoke and nearby shield walls. This field agreed extremely well with the measured field on axis and extends our knowledge of the field off the axis. A Monte Carlo simulation of the experiment with the computed field was compared with the simulation for an ideal field and used to compute the small correction due to the actual field shape.

\subsection{Residual Gas}
Protons that scatter from residual gas molecules can be either neutralized or scattered. Scattered protons will result in a larger time of flight in the wishbone plot. Neutralized protons cannot be detected and so will eliminate some events. Monte Carlo simulations have shown that these two cases tend to produce opposite effects on the measured asymmetry: scattering tends to decrease the asymmetry and neutralization to increase it (due to a slight energy dependence). We accounted for this by performing several data runs with the pressure in the apparatus deliberately raised from about $8\times10^{-5}$ Pa to $3\times10^{-3}$ Pa ($6\times10^{-7}$ Torr to $2\times10^{-5}$ Torr) and then scaling the observed shift in the wishbone asymmetry.

\subsection{Electron Scattering}

Systematic effects due to scatter of beta electrons can be divided into several areas of concern:
\begin{enumerate}
\item Backscatter from the beta spectrometer: Approximately \mbox{5 \%} of electrons incident on the energy detector will backscatter without depositing their full energy,
producing a broad low energy tail in the detector response function. A backscatter veto system was incorporated into the spectrometer design in order to mitigate this effect (see Section \ref{betadet}). The backscatter suppression efficiency was estimated to be about \mbox{90 \%}. Further details can be found in reference \onlinecite{bsNIM}.
\item Scatter from the electron collimator: The collimator was designed with the assistance of a Monte Carlo simulation in order to minimize scattering effects. Thin aperture discs constructed from a high-$Z$ material (tungsten) were employed in order to effectively collimate beta electrons with a minimal surface area available for scattering. The discs were unequally spaced, both to aid collimation and to minimize the number of scattered electrons that subsequently enter the beta spectrometer. The simulation found that \mbox{0.3 \%} of electrons reaching the energy detector would be previously scattered from the collimator. This is comparable to the residual contribution from beta spectrometer backscatter.
\item Bremsstrahlung in the energy detector: There is a small probability (about $10^{-3}$) for an electron to lose energy in the plastic scintillator by radiating a high energy photon that escapes. This also contributes to a low energy tail but it is smaller than the above effects.

\item Scatter from the electrostatic mirror grid wires: Approximately 5 \% of electrons will strike a grid wire when passing through the +3 kV grid above the electron collimator. The typical energy loss is about 100 keV. This effect contributes approximately \mbox{1 \%} to the wishbone asymmetry.

\item Scatter from the top of the vacuum chamber: If a beta electron is emitted in the wrong hemisphere for detection, {\em i.e.} toward the proton detector,  strikes some material, and then backscatters into the beta spectrometer, it will have both the wrong energy and the wrong sign for the asymmetry. This is a particular concern. The apparatus was designed so that all material above the electrostatic mirror, with the exception of the top vacuum flange, is outside the radius of the electron collimator, making it very unlikely for such a backscattered electron to reach the beta spectrometer. The top vacuum flange is coated on the inside with a layer of polyethylene to minimize scattering, and is located above the magnetic flux return in a region of low magnetic field, so the probability of transport back to the beta spectrometer is minimal. As a systematic check, we temporarily replaced the polyethylene layer with a lead sheet to magnify this effect and saw no evidence of it in the wishbone data.

\item Scatter from residual gas: The aCORN operational pressure was $8\times 10^{-5}$ Pa ($6\times 10^{-7}$ Torr) and the residual gas was mostly hydrogen, so electron energy loss is negligible. There is a small probability (approximately \mbox{0.1 \%}) for beta electrons born outside of the aCORN momentum acceptance to be scattered into the beta spectrometer. This does not directly affect the wishbone asymmetry but it does affect the $\phi$ functions (equation \ref{E:phis}) which depend on electron acceptance. The effect is very small (\mbox{$<$ 0.1 \%} of the $a$-coefficient).
\end{enumerate}
Effects 1--3 above contribute to a low energy tail in the detector response function, causing a small fraction (\mbox{$<$1 \%}) of events to be shifted to the left in the wishbone plot (Figure \ref{F:wbMC}). This will tend to fill in the gap between the wishbone branches and can result in an incorrect asymmetry. Our best measure of this came from a search for events in the gap region of the wishbone in the energy region 100 keV to 300 keV where, in the absence of scattering effects, neutron decay events are kinematically forbidden. We found an event rate in that region consistent with zero, but with a \mbox{1 $\sigma$} statistical upper limit (due to background subtraction) corresponding to a \mbox{$<$1.2 \%} low energy tail.

\subsection{Beta Spectrometer Energy Calibration}
The relationship between the measured wishbone asymmetry $X(E)$ and the $a$-coefficient depends on electron energy (equation \ref{E:aEffective}) 
so the energy calibration of the beta spectrometer is important. We use a two-fold strategy. First, two conversion electron sources ($^{207}$Bi
and $^{113}$Bi) are installed {\em in situ} and can be inserted onto the main magnet axis in a gap between the electrostatic mirror and proton
collimator without breaking vacuum.  This is done periodically, at least three times per week, in order to monitor the energy calibration and correct for small variations due to PMT gain drifts. This provides an important relative calibration but we do not rely on this for the final calibration. The absolute energy calibration
is obtained by summing the background subtracted neutron decay wishbone over time to obtain a ``wishbone spectrum'', 
{\em i.e.} the energy spectrum of wishbone events. This is then fit to the theoretical spectrum, essentially the Fermi beta decay spectrum modified
by the aCORN momentum acceptance for coincidence detection of the beta electron and proton. 
From this fit a precise ($<$1 \% relative uncertainty) calibration is obtained. Thus the neutron decay wishbone data are self-calibrating.
Further details can be found in reference \onlinecite{bsNIM}. The neutron decay wishbone spectrum fit is preferred to conversion electron
sources for the absolute calibration because: 1) both the statistical and systematic uncertainties associated with the calibration are smaller; 
2) it is free from source-scattering and energy loss, and the background associated with Compton scattered gamma rays, that complicate the fits to conversion 
electron lines; and 3) it avoids issues related to an apparent rate dependence of the energy pedestal in the data acquisition system that was observed.

\subsection{Proton Collimator Alignment}
A misalignment of the proton collimator is equivalent to a uniform transverse magnetic field. Monte Carlo modeling shows that a 0.1 mrad misalignment would produce a \mbox{0.5 \%} additive error to the $a$-coefficient. In practice the proton collimator was aligned optically to within 0.09 mrad of the magnetic field axis.

\subsection{Proton Scattering}
In an ideal system, any proton that hit a surface in the apparatus would be absorbed and never reach the detector. However there is a small probability for a low energy proton to scatter from a material surface, lose some energy, and still reach the detector with an incorrect time-of-flight. Such events could arise from collisions with the wires of the upper mirror grid, with the walls of the proton collimator, or with one of the proton focusing electrodes (no proton that scattered from the copper walls of the electrostatic mirror could pass through the proton collimator).

A Monte Carlo study using the SRIM\cite{SRIM} ion transport code showed that a proton with energy between 2 keV and 3 keV will be scattered about \mbox{10 \%} of the time with an average loss of about 2/3 of its energy. The majority of these protons will be neutralized to hydrogen and never detected, leaving about \mbox{0.5 \%} of the original decay protons that could be be scattered and then detected. Such scattered protons would produce a tail extending several microseconds beyond the wishbone protons in the time-of-flight plot.

Because the protons are accelerated by the electrostatic mirror, there is a maximum TOF for unscattered protons in the wishbone plot. Scattered neutron decay protons that are not neutralized add a broad tail, several $\mu$s in width, to the TOF response. Our strategy was to look directly for this effect in the wishbone data by summing over electron energy to produce a plot of proton counts against TOF. The total number of protons in the region 1 $\mu$s above the wishbone was compared to the total number in the region 1 $\mu$s below the wishbone and the difference used as a measure of the proton scattering effect. To within statistics, no evidence was found for such scattering events.

\subsection{Proton Focusing}
\begin{figure}[htbp]
\begin{center}
\includegraphics[width = \bc_figure_width]{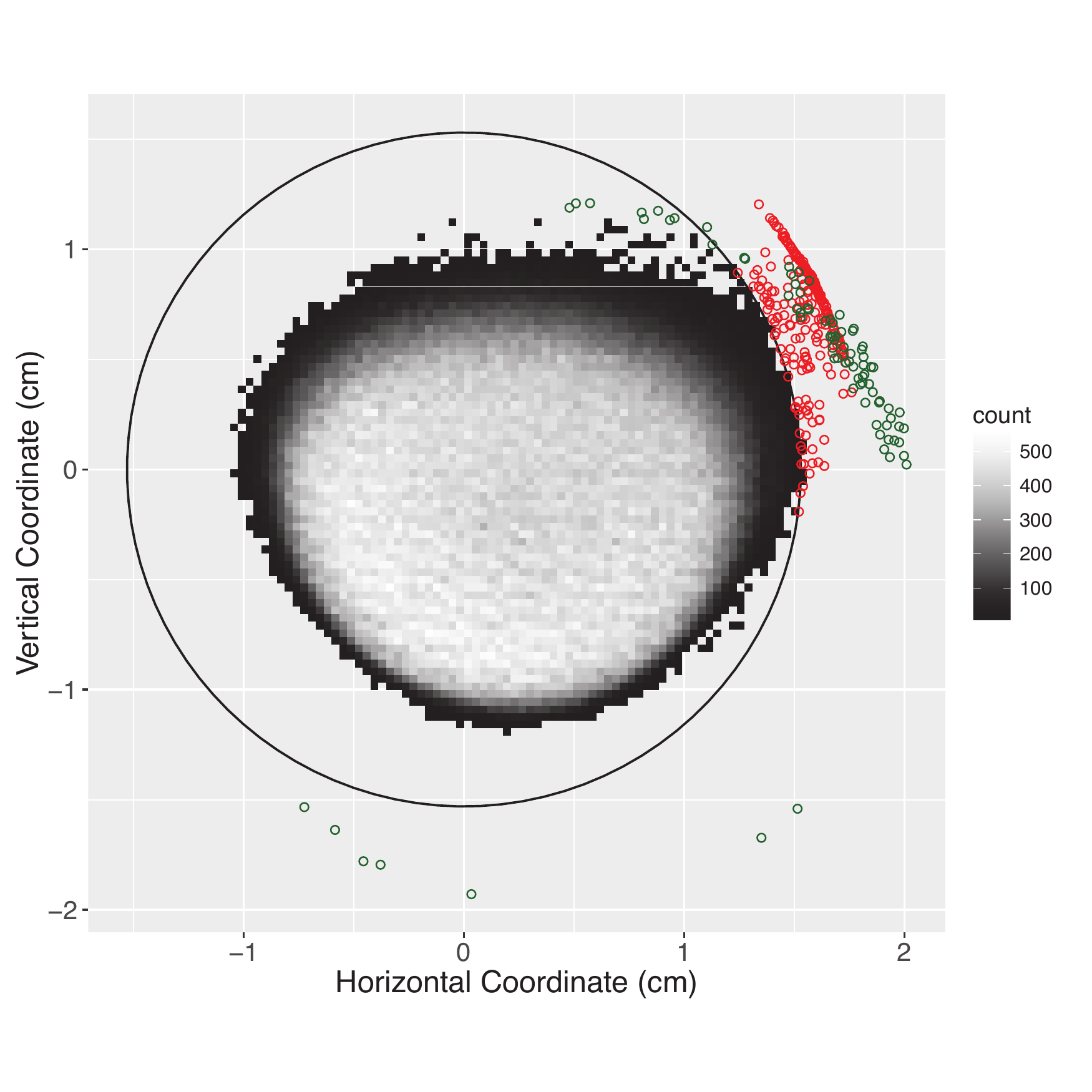}
\end{center}
\vspace{-0.25in}
\caption{\label{F:PSim} Monte Carlo simulation of the distribution of $\approx$one million proton hits on the detector (gray-scale, 99.97 \%), on the detector support (red circles, 0.02 \%), and on the focusing electrodes (green circles, 0.01 \%). The  active area of the detector is shown by the black circle. The elliptic profile of the beam results because the accelerating electric field that guides the protons to the detector is not parallel to the experimental axis.}
\end{figure}

The proton focusing system was designed to accelerate and focus both groups of neutron decay protons onto the detector with high and essentially equal efficiency. A differential efficiency will lead to a false asymmetry. In the experiment we are concerned with imperfect focusing caused by 1) slight mechanical misalignment of components and 2) deviations of the electric and magnetic fields from the design fields. At several
times during the experiment, a FARO\cite{DISCLM,Arm} coordinate measuring machine  was used to locate the three-dimensional positions and  orientations of all electrodes and the detector {\em in situ} relative to the experimental coordinate system. These results were used to make a post-design model of the electric fields for the AMaze simulation. To test the accuracy of the simulated fields compared to the actual fields, 
a set of thin copper masks with different size and shape apertures was made and individually placed in front of the detector. 
For each mask, the ratio of masked to unmasked neutron decay proton events was measured and compared to the equivalent ratios in the simulation. 
By making small adjustments of the proton detector assembly (detector, ring electrode, and fork electrode) position in the simulation, 
good agreement could be found between the simulated mask ratios and the measured ratios. This process produced our best determination of the actual experimental focusing conditions and the uncertainty in position. Figure \ref{F:PSim} shows a high statistics simulation of 10$^{6}$ neutron decay protons transported through from the proton collimator to the detector. Such simulations show that only 0.03 \% of the protons fail to hit the active region of the detector, resulting in a 0.1 \% fractional error in the value of the $a$-coefficient.

\subsection{Neutron Beam Polarization}
If the neutron beam is not completely unpolarized, the antineutrino asymmetry term ($B$-coefficient) in
equation \ref{E:JTWeqn} will introduce a new term that contributes to the wishbone asymmetry
(compare to equation \ref{E:aEffective}, omitting the small corrections):
\begin{equation}
X(E) = a f_a(E) \pm P B f_B(E)
\end{equation}
where $P$ is the neutron polarization and $f_B(E)$ is a geometric acceptance function associated with the
antineutrino asymmetry. The positive (negative) sign applies when the axial magnetic field direction 
is up, toward the proton detector (down, toward the electron detector). The fact that the $B$-coefficient
is about ten times larger than the $a$-coefficient, and that a phase space enhancement makes $f_B(E)$
about 40 \% larger than $f_a(E)$, together make the experiment very sensitive to neutron polarization. The NG-6 beam is, in principle, unpolarized, but the neutron guide wall is $^{58}$Ni (magnetic) and the presence of superconducting magnetics in its vicinity, past and present, makes a slight unwanted neutron polarization possible. Unfortunately we were unable to directly measure the neutron polarization on NG-6.
\par
We collected data with both directions of the axial magnetic field. A simple average of the 
$a$-coefficients obtained with magnetic field up ($a_{\rm up}$) and down ($a_{\rm down}$) cancels the polarization effect, assuming that $P_{\rm up} = P_{\rm down}$. There is an additional correction in the case of a small difference in polarization:
\begin{equation}
\overline{a} = \frac{1}{2}\left( a_{\rm up} + a_{\rm down}\right) + 
 \frac{1}{2}\left( \frac{P_{\rm down} - P_{\rm up}}{P_{\rm down} + P_{\rm up}}\right) \left( a_{\rm up} - a_{\rm down}\right)
\end{equation}

\section{Conclusion}
The aCORN apparatus operated on the NIST NG-6 beamline from February 2013 through May 2014, collecting 1900 hours of data. These data have been analyzed and will be the subject of a forthcoming paper. The apparatus was then moved to the higher flux NG-C beamline where an improved electrostatic mirror was installed and the main magnet reconfigured to improve the magnetic field at the top of the proton collimator. The apparatus collected data on the NG-C beamline from mid 2015 to late 2016 before being placed into storage. For the future, we are investigating the possibility of performing an experiment with the same apparatus and a polarized neutron beam to measure the antineutrino asymmetry coefficient, $B$.

\begin{acknowledgments}
The authors acknowledge the management and staff of the NIST Center for Neutron Research for providing the neutron facilities used in this work. Technical support was provided by Jeremy Cook, Daniel Adler, and Eli Baltic at the NCNR and by Walter Fox, Mark Leuschner, Gerard Visser, Tom Rinckel, John Vanderwerp, and Jack Doskow at CEEM. 
We would also like to thank a number of students for their work over the life of this project; Tomasz Konopka, Corbin Butcher, Glenn Smith, Andrew Portugese, and Edward Lamere and others from Hamilton College, Melanie Novak and Rana Ashkar from Indiana University, and Sarang Wardadkar, Shelby Vorndran, Michelle Whitehead, Boyu Meng, Chris Clark, and Kaitrin Higbee from DePauw University.

This work was supported by the National Science Foundation, U.S. Department of Energy Office of Science, and NIST (US Department of Commerce). We 	thank the NIST Center for Neutron Research for technical support and for providing the neutron facilities used in this work. 
\end{acknowledgments}


\begin{thebibliography}{99}
\bibitem{DISCLM} Certain trade names and company products are 
mentioned in the text or identified in illustrations in order to 
adequately specify the experimental procedure and equipment used. In 
no case does such identification imply recommendation or endorsement 
by the National Institute of Standards and Technology, nor does it imply 
that the products are necessarily the best available for the purpose.
\bibitem{Bearings} Bearings are radial ball bearings from SKF. Top bearing is model 16019 with 95~mm bore. Bottom bearing is model 16034 with 170~mm bore.
\end{thebibliography}
\end{document}